\documentclass{aa}  

\usepackage{placeins}
\usepackage{natbib,twoopt}
\usepackage[breaklinks=true]{hyperref} %% to avoid \citeads line fills
\bibpunct{(}{)}{;}{a}{}{,}             %% natbib format for A&A and ApJ
\usepackage{graphicx}
\usepackage{txfonts}
\usepackage{hyperref}
\usepackage{booktabs}
\hypersetup{
    colorlinks=true,
    linkcolor=blue,
    citecolor = blue,
    filecolor=magenta,      
    urlcolor=blue}
\def\kms{km~s$^{-1}$}

\graphicspath{{Figures/}}

\authorrunning{Bellazzini et al.}
\titlerunning{Ultra-faint compact satellites of the Milky Way. The case of Koposov~2}

\newcommand{\Xo}    {X_0}
\newcommand{\Yo}    {Y_0}
\newcommand{\Rc}    {R_{\rm c}}
\newcommand{\Rh}    {R_{\rm h}}
\newcommand{\rh}    {r_{\rm h}}
\newcommand{\Rt}    {R_{\rm t}}
\newcommand{\RJ}    {r_{\rm J}}
\newcommand{\Sigmao}{\Sigma_0}
\newcommand{\PP}    {\mathcal{P}}
\newcommand{\btheta}{\boldsymbol{\Theta}}
\newcommand{\Area}  {\mathcal{A}}
\newcommand{\logten}{\log_{10}}
\newcommand{\ASigma}{A_{\Sigma}}

\newcommand{\Xj}    {X_i}
\newcommand{\Yj}    {Y_i}

\newcommand{\dd}    {{\rm d}}
\newcommand{\LL}    {\mathcal{L}}
\newcommand{\DD}    {\mathcal{D}}

\newcommand{\UU}    {\mathcal{U}}
\newcommand{\NN}    {\mathcal{N}}

\begin{document}

\title{The {\em Hubble} Missing Globular Cluster Survey}
\subtitle{IV. Ultra-faint compact satellites of the Milky Way: the case of Koposov~2}

\author{M.~Bellazzini\inst{1}, R.~Pascale\inst{1}, 
E.~Dodd\inst{2}, E.~Ceccarelli\inst{3}\fnmsep\inst{1}\fnmsep\inst{4}, 
D.~Massari\inst{1}, M.~Libralato\inst{5}, 
S.~Cassisi\inst{6,7}, M.~De Leo\inst{4}, A.~Della Croce\inst{8}, E.~Vesperini\inst{8},
A.~Mucciarelli\inst{4}\fnmsep\inst{1}, E.~Dalessandro \inst{1}, M.~Salaris\inst{1}, F.~Aguado-Agelet\inst{9,10}, A.~Bellini\inst{11}, F.R.~Ferraro\inst{4}, B. Lanzoni\inst{4},  M.~Monelli\inst{6,12,13}, S. Saracino\inst{14}
}

\institute{
INAF, Osservatorio di Astrofisica e Scienza dello Spazio di Bologna, Via Gobetti 93/3, I-40129 Bologna, Italy
   \email{michele.bellazzini@inaf.it}
   \and
Institute for Computational Cosmology \& Centre for Extragalactic Astronomy, Department of Physics, Durham University, South Road, Durham, DH1 3LE, UK
\and
Kapteyn Astronomical Institute, University of Groningen, Landleven 12, 9747 AD Groningen, The Netherlands
\and
Dipartimento di Fisica e Astronomia, Universit\`a  degli Studi di Bologna, Via Gobetti 93/2, I-40129 Bologna, Italy
\and
INAF - Osservatorio Astronomico di Padova, Vicolo dell'Osservatorio 5, Padova 35122, Italy
 \and
             INAF – Osservatorio Astronomico di Abruzzo, Via M. Maggini, 64100 Teramo, Italy
          \and
             INFN, Sezione di Pisa, Largo Pontecorvo 3, 56127 Pisa, Italy
\and
Department of Astronomy, Indiana University, Swain West, 727 E. 3rd Street, IN 47405 Bloomington, USA
\and
atlanTTic, Universidade de Vigo, Escola de Enxe\~nar\'ia de Telecomunicaci\'on, 36310, Vigo, Spain
\and
Universidad de La Laguna, Avda. Astrof\'isico Fco. S\'anchez, E-38205 La Laguna, Tenerife, Spain
\and
Space Telescope Science Institute, 3700 San Martin Drive, Baltimore, MD 21218, USA
\and
IAC- Instituto de Astrof\'isica de Canarias, Calle V\'ia Lactea s/n, E-38205 La Laguna, Tenerife, Spain
         \and
             Departmento de Astrof\'isica, Universidad de La Laguna, E-38206 La Laguna, Tenerife, Spain
\and
INAF - Osservatorio Astrofisico di Arcetri, Largo E. Fermi 5, 50125 Firenze, Italy
}
 
\abstract
{In the last decades a number of extremely faint  
and compact Galactic satellites (Ultra Faint Compact Satellites; UFCS) have been discovered by large panoramic surveys. The nature of these stellar systems is uncertain, both because they occupy a region of the $M_V-\Rh$ plane where dwarf galaxies and star clusters overlap, and for the observational challenges inherent with their faintness and distance. Here we show how the deep HST photometry from the Missing Globular Clusters Survey (MGCS) can provide new insight into the nature of these satellites, especially when coupled with spectroscopic metallicity measures, by providing unrivalled accurate estimates of their distance and age. As an example, we consider the intriguing case of Koposov~2, currently the most metal-poor bound star cluster known in the entire Milky Way or an extreme case of Ultra Faint Dwarf galaxy. By performing a spectroscopically-informed bayesian isochrone fit on the MGCS data we find $(m-M)_0=16.85\pm 0.06$ ($D=23.4\pm 0.6$~kpc) and age=$13.7^{+0.9}_{-1.3}$~Gyr, showing that, contrary to previous age estimates, Koposov~2 is as old as the oldest Galactic globular clusters. The luminosity function, corrected for incompleteness, is well reproduced by a model with the same age and metallicity and a slope of the mass function $x=-0.35$, suggesting a significant depletion of faint stars. 

We also model the surface stellar density field, deriving new robust estimates of the half-light radius ($\Rh=0.39^{+0.06}_{-0.04}$~arcmin, corresponding to $\Rh=2.7^{+0.4}_{-0.3}$~pc), of the absolute integrated magnitude ($M_V=-0.95 \pm 0.22$) and of the stellar mass ($M_{\star}=371.8\pm 41.6~M_{\sun}$), showing that Koposov~2 is much more compact than { confirmed} dwarf galaxies of similar stellar mass. In summary, the new observational evidence provide significant support to the hypothesis that Koposov~2 is a star cluster that may have lost a large fraction of its original mass.
Finally we show that most UFCS lie in the same locus of the $M_{\star} - \Rh$ plane as Galactic open clusters, hinting to a possible additional channel for their formation. 

}

   \maketitle
%
%-------------------------------------------------------------------
\section{Introduction}
\label{sect:intro}

In the last decades, panoramic surveys performed with CCD cameras probed an unexplored range of the parameter space of stellar aggregates, leading to the discovery of several new faint satellites of the Milky Way \citep[MW; see, e.g.,][and references therein]{simon_19,pace_25,geha_26}. The newly discovered objects blurred the distinction between the different classes of low-mass stellar systems, forcing a re-consideration of the very definition of the terms "galaxy" and "star cluster" \citep{ws12}.

The faintest and the most compact among them are probably the most difficult to classify as, (a) they lie in a region of the classical luminosity - size plane that is between the loci of dwarf galaxies and of globular clusters (GCs), (b) their intrinsic faintness implies that their Red Giant Branch (RGB) stars are very rare or non-existing. This, coupled with their relatively large distance ($D> 10$~kpc) makes their stars challenging spectroscopic targets. 

%--------------------
\begin{figure}[h!]
\centering
\includegraphics[width=\columnwidth]{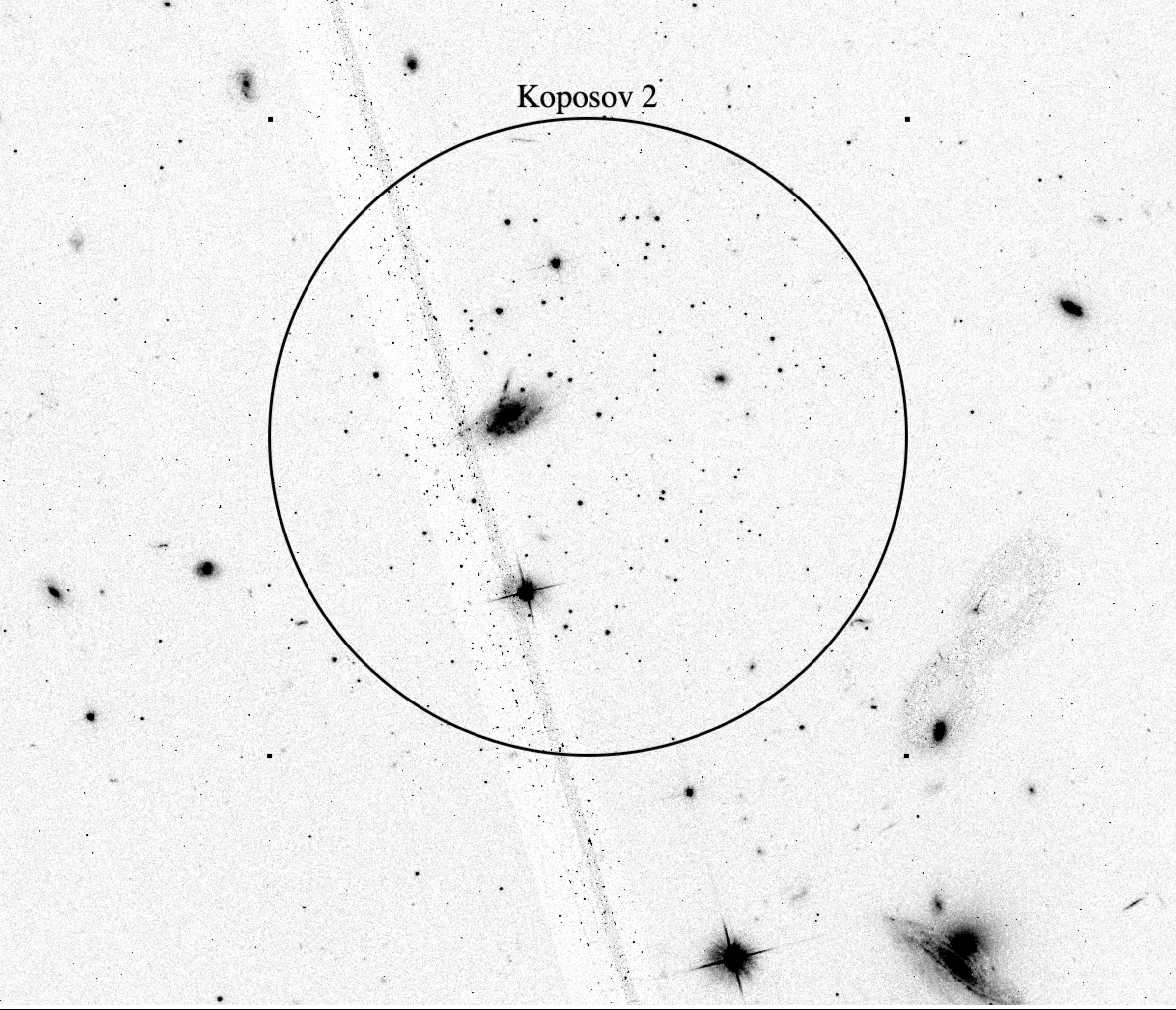}
\caption{Central portion of the FoV of our HST ACS/WFC F606W stacked image of Kop~2. North is up, East to the left and the black circle is centred on the center of the cluster and it has a radius of $30.0\arcsec$, corresponding to $ \simeq 1.3~\Rh$. 
The narrow diagonal band crossing the image corresponds to the inter-chip gap of the ACS/WFC and the eight-shaped feature in the lower right quadrant is due to internal reflections of the telescope, likely due to a nearby bright foreground star not shown here.} \label{fig:ima}
\end{figure}
%--------------------

In a recent paper collecting and analysing existing spectroscopic data on these systems, \citet[][C26 hereafter]{cerny_26} proposed an empirical definition of this class, denoted as Ultra Faint Compact Satellites (UFCS), as stellar systems { lying at significant height above the Galactic plane ($\rm |Z|>5.0$~kpc)}, having absolute magnitude $M_V>-3.5$, average surface brightness within the half-light radius $\mu_{V,\,\Rh}>24.0~ \rm mag~arcsec^{-2}$ and linear half-light radius $\Rh \lesssim  15.0$~pc. C26 considered 19 Galactic UFCS and only in six cases they were able to provide a robust classification as "{ definite star cluster / likely or very likely star cluster}" or "{ likely} dwarf galaxy" 
on the basis of the six discriminating criteria they adopt, namely age (assuming that an UFCS younger than $\simeq 10.0$~Gyr should be a star cluster, not an Ultra Faint Dwarf galaxy - UFD), mean metallicity (adopting the limit $\rm [Fe/H]\gtrsim -2.5$ for star clusters, while UFDs are typically more metal-poor), velocity dispersion ($\sigma_{V}$\footnote{While, in principle, the dispersion in any component of the velocity could be used as a diagnostic, in practice for these systems only the dispersion of the velocity along the line of sight is within reach. For this reason, in the following we will always refer to $\sigma_{V_{\rm los}}$, when appropriate.} requiring the presence of a dark matter halo would imply that the system is a galaxy), metallicity spread \citep[a sizeable $\sigma_{\rm [Fe/H]}$ is a distinctive feature of galaxies vs. star clusters;][]{ws12}, the presence of stellar mass segregation \citep[that would favour a classification as a star cluster, but see, e.g.][]{errani25,errani26}, and finally the presence of carbon-enhanced metal-poor stars with no
neutron-capture element enhancement (CEMP-no stars) that would favour classification as UFD \citep[see,][and references therein]{simon24,kotsou26}. It is important to note that only two of the above criteria can provide clear-cut answers on the nature of UFCS, those based on the parameters that are the most challenging to measure in these systems, $\sigma_{V}$ and $\sigma_{\rm [Fe/H]}$ \citep[C26,][]{geha26_I,geha_26,ceccarelli_26}, while the others rely, e.g., on educated guesses on properties of extremely metal-poor star clusters of which we have no actual examples \citep[but see][]{iba24}.

One of the most intriguing and ambiguous cases studied by C26 is the UFCS Koposov~2 \citep[Kop~2 hereafter,][]{koposov_07,paust14,munoz18}, whose velocity dispersion is barely resolved and highly uncertain. No measure of its $\sigma_{\rm [Fe/H]}$ is available, but a very likely member star near the base of the RGB was found to have $\rm [Fe/H]\simeq -2.9$ \citep[C26;][]{geha26_I, ceccarelli_26}, thus leaving open the options of an extremely faint UFD \citep[$M_V=-0.92\pm 0.81$][]{munoz18} or of the most metal-poor bound star cluster ever observed. The large uncertainty in the distance and age of Kop~2 is a major problem in ascertaining its nature (C26). All the available estimates are based on fitting the Turn Off (TO) and the upper Main Sequence (MS) of Colour Magnitude Diagrams (CMD) obtained from ground-based observations with theoretical isochrones, and adopting a fixed metallicity. For example, by assuming [Fe/H]=-2.0, \citet{koposov_07} derived age=8~Gyr and $D\simeq 40$~kpc. 
while, assuming [Fe/H]=-0.6, \citet{paust14} derived age=4-6~Gyr and $D=33.5\pm 1.5$~kpc. Finally, C26 noted that they were able to obtain a more consistent target selection for spectroscopic follow-up by adopting an isochrone with [Fe/H]=-2.2 and age=13.5~Gyr shifted to $D\simeq 24$~kpc. However, C26 recognised that it is hard to achieve a fully satisfactory fit of existing CMDs of this cluster \citep[see, e.g., Fig.~8 of][]{paust14}, tentatively blaming a possible contamination from the Sgr dwarf spheroidal galaxy (hereafter dSph) tidal stream \citep{iba20,ramos22} projected onto the cluster background. We note that a possible association of Kop~2 with the tidal stream of the Sgr~dSph galaxy, suggested by a match in projection on the celestial sphere \citep{paust14,mic_sgr}, has been refuted by \citet{ceccarelli_26} on the basis of 3D kinematics. We independently checked that the motion of the system does not match the predictions of the model of the Sgr stream by \citet{vasi21}, also given the new distance derived in the present paper.

In this paper, we take Kop~2 as an example of how the deep Hubble Space Telescope (HST) photometry from the {\em Hubble Missing Globular Clusters Survey} 
\citep[MGCS;][]{massari_25,libra26} can aid in understanding the nature of UFCS, by providing distance and age estimates with unmatched accuracy, thanks to the spatial resolution and photometric precision of HST data processed by the MGCS team. 
These data enabled us also to derive new, more robust and accurate estimates of structural parameters and integrated magnitude of the system and to get deeper insight into the luminosity and mass function, as well as into the degree of stellar mass segregation. There are several UFCS among the MGCS targets for which the same kind of analysis can be performed.

In Sect.~\ref{sect:cmd} we briefly present the data and discuss the CMD of the cluster, as well as its contamination from back-/foreground sources. In Sect.~\ref{sect:agedist} we measure the age, distance and interstellar extinction of Kop~2 using the Bayesian framework developed within the Cluster Ages to Reconstruct the Milky Way Assembly (CARMA) collaboration \citep[][]{massari23}, that has already proven to be able to provide sub-Gyr age precision on a large sample of MW \citep[][]{aguado_agelet_25, massari26} and Magellanic Clouds \citep[][]{niederhofer25} star clusters.  In Sect.~\ref{sect:struc} we use again a Bayesian approach to derive the structural parameters of the system, taking into account all the relevant selection effects.  In Sect.~\ref{sec:lf}, we compare the observed luminosity function (LF) with theoretical models, constraining the slope of the stellar mass function (MF), and we derive the integrated absolute magnitude of the system. 
We also test the presence of stellar mass segregation.
Finally, In Sect.~\ref{sect:disc} we put Kop~2 and the UFCS population in context and we discuss our findings and, finally, in Sect.~\ref{sec:conclu} we summarise our results and draw our conclusions. In Appendix~\ref{app:cf} we describe the artificial star experiments by which we determine the Completeness Fraction (CF) as a function of magnitude, in Appendix~\ref{app:isofit} we provide additional material on the isochrone fitting process, and in Appendix~\ref{app:orb} and we describe how we re-derive the orbit of Kop~2 within the MW gravitational potential in light of our newly derived distance and of the systemic motion by \citet{libra26} and \citet{ceccarelli_26}. We consider the effect on the derived orbit of different Galactic potentials and of the Large Magellanic Cloud (LMC). In Appendix~\ref{app:evol} we present simple $N$-body models of the evolution of  star clusters in a weak tidal field whose end state is similar to Kop~2.

%-------------------------------------------------------------
\begin{figure*}
\centering
\includegraphics[width=0.9\textwidth]{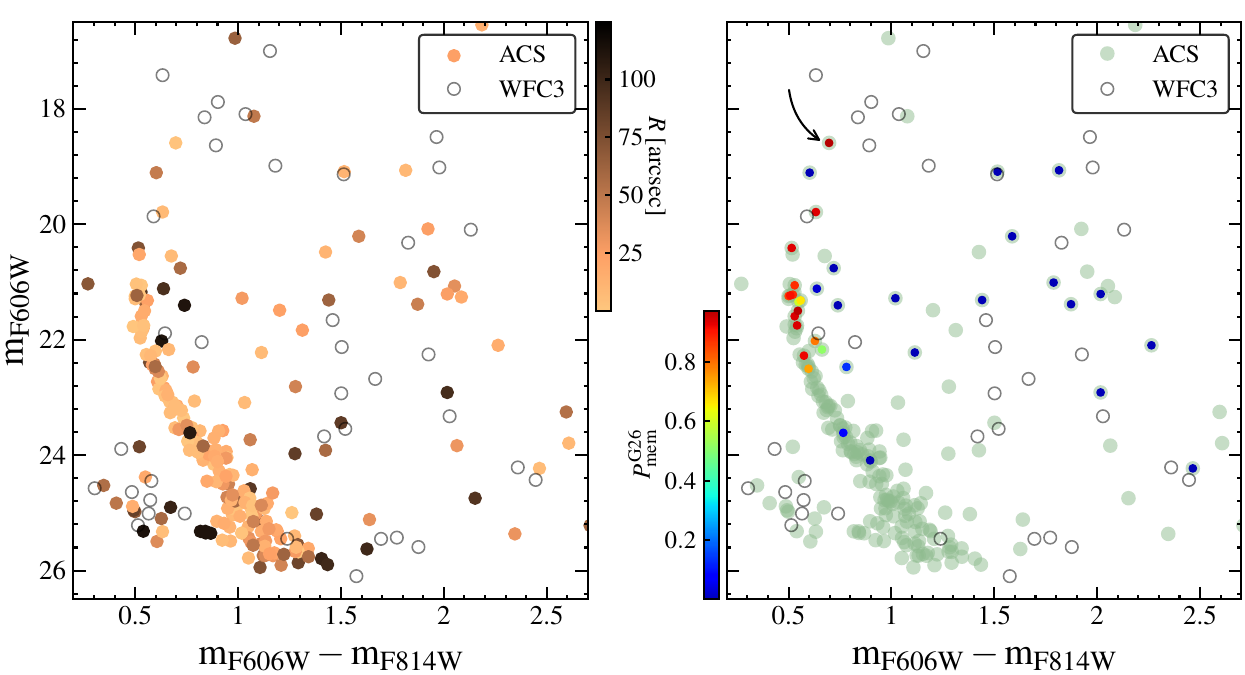}
\caption{CMDs of stars in the ACS/WFC field (filled coloured circles) and of stars in the parallel WFC3/UVIS field (black empty circles). In the left panel ACS stars are colour-coded according to their angular distance from the center of Kop~2. In the right panel we marked the stars in the ACS/WFC field for which \citet{geha26_I} measured line-of-sight velocities as small filled circles colour-coded according to the probability of membership to Kop~2 after the same authors. The arrow in the right panel indicates the RGB star for which metallicity is available.}
\label{fig:CMD}
\end{figure*}
%-------------------------------------------------------------

%-------------------------------------------------------------------
\section{The data and the colour magnitude diagram} \label{sect:cmd}

HST MGCS data for clusters not affected by high extinction, including Kop~2, are taken with the wide field channel (WFC) of Advanced Camera for Survey (ACS) for a field approximately centred on the target, plus parallel observations with the ultraviolet and visible channel (UVIS) of Wide Field Camera 3 (WFC3). In the case of Kop~2 the parallel WFC3 field is located $\simeq 5.0$ arcmin from the center of the cluster. The tidal radius of Kop~2 is $\simeq 2.0$ arcmin (Sect.~\ref{sect:struc}), and thus the cluster is almost entirely enclosed in the $3.4\,{\rm arcmin} \times 3.4\,{\rm arcmin}$ ACS-WFC field of view and the WFC3/UVIS parallel field samples only the fore/background populations.
All the data of MGCS clusters observed with the ACS-WFC and WFC3-UVIS have been acquired using the F606W and F814W filters, and adopting the same dithering scheme and combination of deep and long exposures described in detail in \citep{massari_25,libra26}. Photometry and astrometry have been obtained with standard state-of-the art Point Spread Function (PSF) fitting technique, as described in \citet{libra26}, where the artificial-stars experiments that have been performed to probe the completeness of the sample as a function of magnitude, colour and position in the frames are also described. In Fig.~\ref{fig:ima} we show the central part of the drizzled ACS-WFC image centred on Kop~2, showing how sparse and faint the stars of this system are, and how remarkably uncrowded Kop~2 is as seen with HST. The narrow diagonal stripe that appears to cross the image is the trace of the inter-chip gap. This region and the outer margins of the two ACS-WFC chips were exposed for less time than the rest of the chip because of the dither pattern. These small sub-exposed regions are therefore excluded in any subsequent analysis based on star counts. 

In Fig.~\ref{fig:CMD} we show two versions of the MGCS CMD of Kop~2. Following \citet{massari_25} and \citet{rosignoli26}, we show here only stars selected according to the magnitude-dependent criteria based on the quality parameters provided in the MGCS catalogues, as described in \citet{libra22, libra26}. { Saturation begins to affect the clean measure of stellar flux for $\rm m_{F606W}<19.4$, but the adopted photometry code is able to successfully recover the magnitudes of partially saturated stars with typical precision of about $\simeq 0.02$~mag \citep[from HST ACS-WFC data;][]{anderson22, nardiello18}.} In the left panel of Fig.~\ref{fig:CMD} the stars from the ACS fields are colour-coded according to their distance from the cluster center, while the stars from the parallel $2.7\,{\rm arcmin} \times 2.7\,{\rm arcmin}$ WFC3 field are plotted as empty black circles\footnote{The ACS and the WFC3 fields are affected by indistinguishable amounts of interstellar extinction. According to the \citet{sfd98} maps, recalibrated following \citet{schlafly2011}, the former has $\rm E(B-V)=0.037 \pm 0.002$ and the latter $\rm E(B-V)= 0.041\pm 0.002$.}. The narrow and very well-defined MS of Kop~2, ranging from the TO point, around $\rm m_{F606W}=21.0$ and $\rm m_{F606W}-m_{F814W}\simeq 0.6$ to the limiting magnitude $\rm m_{F606W}\simeq 26.0$, where it reaches $\rm m_{F606W}-m_{F814W}\simeq 1.2$, is clearly dominated by stars within $\simeq 60\,{\rm arcsec}$ from the cluster centre. The comparison with the distribution of the stars in the WFC3 field shows that the contamination of the cluster MS by fore/background sources, either from the MW or the Sgr Stream, is negligible. It is interesting to note that the curious grouping of stars seen in both the ACS and WFC3 fields around $\rm m_{F606W}-m_{F814W}\simeq 0.5$ and $\rm m_{F606W}=25.0$, to the blue of the MS of Kop~2, is consistent with the upper MS of an age-intermediate population ($\simeq 8$~Gyr) at the distance where the present day snapshot of the model of the disruption of the Sgr dSph by \citet{vasi21} locates the densest branch of the Sgr Stream in the background of the cluster ($D\simeq 120$~kpc). Another remarkable feature of this CMD is the lack of a pronounced sequence of nearly equal mass binaries running parallel to the red of the MS, which is quite typical in such sparse clusters and that is clearly evident in the MGCS CMDs of similar clusters. 

In the right panel of Fig.~\ref{fig:CMD} we show the same CMD 
with the stars for which \citet{geha26_I} provides line-of-sight velocity measures that are highlighted with a small filled circled, colour-coded according to the probability membership computed by the same authors ($P_{\rm mem}^{\rm G26}$)\footnote{ While $P_{\rm mem}^{\rm G26}$ is computed combining information from kinematics, location in the CMD, metallicity, when available, as well as other parameters, it must be noted that, in this case, all the stars with $P_{\rm mem}^{\rm G26}>0.6$ have line-of-sight velocity within 3$\sigma$ of the systemic velocity of Kop~2, as derived by C26, with only one exception whose velocity is within 5$\sigma$. Hence, high $P_{\rm mem}^{\rm G26}$ stars should be considered as high-likelihood radial velocity members.}.
The spectroscopic members not only strongly support the observed morphology of the TO region, but identify two likely RGB stars, around $\rm m_{F606W}-m_{F814W}\simeq 0.7$, between $\rm m_{F606W}=18.0$ and $\rm m_{F606W}=20.0$.

For the brightest of the two RGB members (Gaia DR3 $874275674394169984$), there are now two fully independent and fully compatible spectroscopic metallicity measurements, both based on the Ca triplet around $8600 \AA$. These are $\rm [Fe/H]=-2.91\pm 0.12$ by \citet{ceccarelli_26}, and $\rm [Fe/H]=-2.89\pm 0.18$ by \citet{geha26_I}, while C26, from the same data as \citet{geha26_I} reports $\rm [Fe/H]=-2.86\pm 0.18$\footnote{In this context it may be interesting to note that \citet{geha26_I} reports a similar metallicity also for the second RGB member, albeit with a very large uncertainty $\rm [Fe/H]=-2.85\pm 0.69$.}. There is little doubt that this star is indeed an actual member of the cluster. The proper motion and the radial velocity are compatible with the systemic motion \citep[C26;][]{ceccarelli_26}\footnote{It may be worth noting that the Gaia DR3 parallax for this star is $\pi=0.287 \pm 0.198$~mas. While fully compatible with $\pi\simeq 0.0$ (within 1.45$\sigma$), this parameter deserve additional check in future Gaia releases to verify compatibility with membership to such a distant satellite as Kop~2.}. The occurrence of an interloper from the Sgr dSph with such a low metallicity is highly unlikely (C26). Moreover, as anticipated above, the motions of stars in the branches of the Sgr Stream in the foreground (at $D\sim 15$~kpc) and in the background (at $D\sim 80-120$~kpc) is predicted to be significantly different from that of Kop~2 \citep{vasi21}. Finally, \citet{ceccarelli_26} concluded that also a Galactic interloper is very unlikely, based on simulations performed with the Besancon Galactic model\footnote{\url{https://model.obs-besancon.fr}} \citep{robin12}.
In the following analysis, we will adopt the metallicity of this star as a robust prior for our age and distance determination. However, it is worth noting that even when largely uninformative metallicity priors were adopted, our Bayesian procedure of isochrone fitting invariably converged toward very low metallicity values.

\section{Age and distance} \label{sect:agedist}

%-------------------------------------------------------------
\begin{figure*}[h!]
\centering
\begin{minipage}{0.45\textwidth}
\centering        
\includegraphics[width=0.95\textwidth]{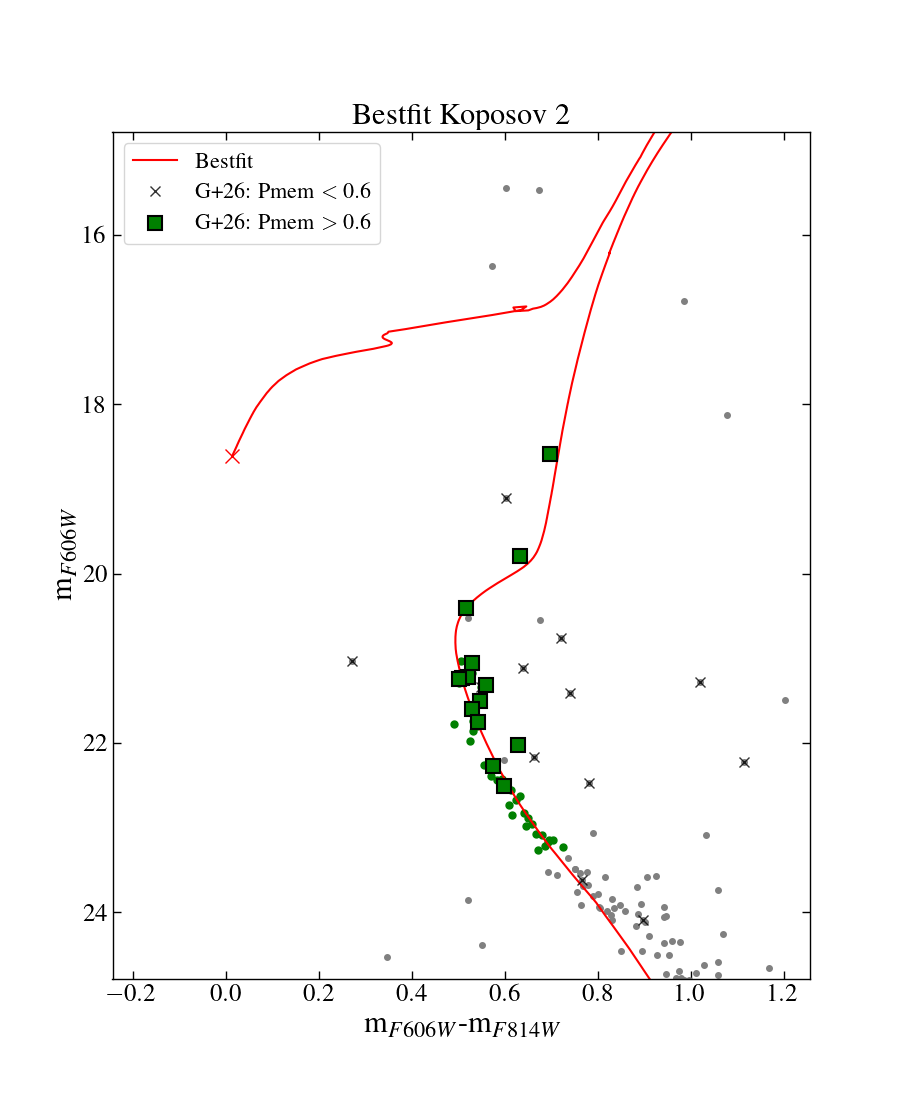}           
\end{minipage}
\begin{minipage}{0.45\textwidth}
\centering
\includegraphics[width=1.0\textwidth]{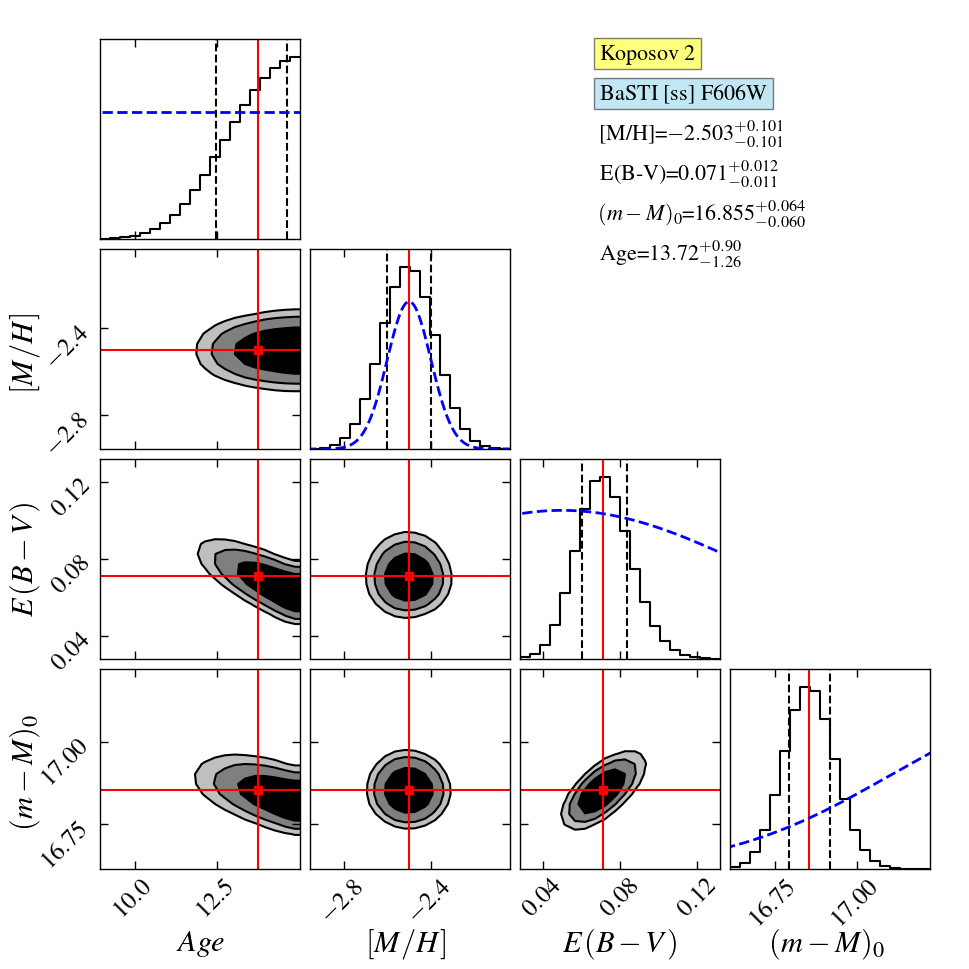}    
\end{minipage}  
\caption{Results of the Bayesian isochrone fit in the ${\rm m}_{\rm F606W}-{\rm m}_{\rm F814W}$ vs. ${\rm m}_{\rm F606W}$ plane. In the left panel the best-fit isochrone is superimposed to the observed CMD after correcting for the estimated distance modulus and colour excess. The stars used for the fit are plotted in green, with square markers and a black outline highlighting radial velocity members according to \citet{geha26_I} and non-members marked with crosses. In the right panel the marginalised two- and one-dimensional posterior distributions of the model free parameters are shown.
The adopted prior PDFs (dashed blue lines) are plotted together to the posterior PDF (black histograms) in the one-dimensional posterior distributions. In these sub-panels the red vertical line marks the median of the posterior PDF and the dashed black lines enclose the range from the 16th to the 84th percentile of the distribution. In the 2D distributions the red crosses and the red dot marks the median of the posterior PDFs.   }
\label{fig:agedist}
\end{figure*}
%-------------------------------------------------------------

To get the best measure of the age and distance of Kop~2 from isochrone fitting of our narrow cluster MS, we use the Bayesian procedure adopted by the CARMA project \citep{massari23,aguado_agelet_25}, that is also the standard adopted by MGCS \citep{massari_25}. 
The algorithm uses Markov Chain Monte Carlo sampling \citep[as implemented in  the {\tt emcee} software library,][]{emcee} to explore the  posterior distribution of a model defined by four parameters, namely age, distance modulus, E(B-V), and [M/H]. The likelihood is evaluated by comparing the observed distribution of cluster stars in the CMD with a fine grid of BaSTI-IAC\footnote{\url{http://basti-iac.oa-abruzzo.inaf.it}} isochrones \citep[][and references therein]{basti18,basti21}.

Here [M/H] is the global metallicity, including the contribution of all the chemical elements heavier than Helium. This is a straightforward approach to incorporate the effects of the iron and $\alpha$-elements abundance into a single parameter. In general,  an isochrone of given age and given $\rm [Fe/H]$ and $\rm [\alpha/Fe]>0.0$ is indistinguishable from an isochrone of the same age with solar-scaled abundance having
\begin{equation}
    {\rm [M/H]} = {\rm [Fe/H]} + \log(0.694\times10^{[\alpha/{\rm Fe}]} +0.306)
\end{equation}
\label{eq:mh}
as demonstrated by \citet{scs93} and \citet{cassisi04}\footnote{The coefficients of Eq.~\ref{eq:mh} are slightly different from those of the original formula by \citet{scs93} to take into account the use of a different reference solar mixture. The BaSTI models are based on the solar heavy-element distribution provided by \citet{caffau_11}, whereas the models by \citet{scs93} were based on the \citet{ross76} mixture.}.
The adopted isochrone grid span the metallicity range $\rm -3.0\le [M/H]\le 0.0$ with 0.01~dex step, and the age range from 1.0~Gyr to 15.0~Gyr in steps of 0.1~Gyr.    

In the left panel of Fig.~\ref{fig:agedist} we show the best-fit isochrone superposed to the observed CMD of Kop~2 with the \citet{geha26_I} spectroscopic targets with $P^{\rm G26}_{\rm mem}>0.6$ using square markers, while in the right panel we show the corner-plot of the corresponding fit. 
For $\rm m_{F606W}<21.0$ only stars  with $P^{\rm G26}_{\rm mem}>0.6$ have been used for the fit.
In the sub-panels of the corner plot showing the histograms of the marginalised posterior PDF of each individual parameter, the adopted prior distribution are also shown as blue dashed curves. The prior on age is uniform over the entire age grid, that on E(B-V) is a very broad gaussian centred on the value obtained from the \citet{sfd98} maps recalibrated by \citet{schlafly2011}, also adopted by \citet{paust14}, and the prior on distance is also a broad gaussian centred on the distance estimate by \citet{paust14}. The most informative prior is that on [M/H], where we adopted a gaussian distribution with mean $\rm [M/H]=-2.5$ and $\sigma_{\rm [M/H]}=0.1$, taking the spectroscopic $\rm [Fe/H]=-2.9$ value for reference and hypothesising $\rm [\alpha/Fe]\simeq +0.5$ \citep{euge26}. More details on the adopted priors are provided in Appendix~\ref{app:isofit}, where we also show that adopting a prior on metallicity corresponding to  $\rm [Fe/H]=-2.9$ and $\rm [\alpha/Fe]\simeq +0.3$ has a negligible effect on the analysis. 

The solution shown in Fig.~\ref{fig:agedist} provides a strong constraint on the distance modulus, $\rm (m-M)_0=16.85\pm 0.06$, corresponding to $D=23.4\pm 0.6$~kpc, in good agreement with that hypothesised by C26 and much closer than the distances estimated by \citet{koposov_07}, $D\sim 40$~kpc, and by \citet{paust14}, $D= 33.3\pm 1.5$~kpc. Very old ages are also strongly preferred, with the mode of the PDF lying at the extreme of the range spanned by the underlying grid of models. It is worth stressing here that we have repeated the fit adopting a wide range of priors and in all the cases the solution converges toward the same distance modulus and old age, within the uncertainties. The same is true for $\rm E(B-V)=0.07\pm 0.01$, that is only slightly larger than  $\rm E(B-V)=0.037$ as read from the \citet{sfd98} low-spatial-resolution maps recalibrated by \citet{schlafly2011}. The metallicity PDF is virtually indistinguishable from the prior. We note that the very old age we infer for Kop~2, is significantly different from the estimates by \citet[][$\sim 8$~Gyr]{koposov_07} and by \citet[][$\sim 5$~Gyr]{paust14}, and much more consistent with the very low spectroscopic metallicity. Following the standard prescriptions by the CARMA and MGCS projects, we take as our best fit estimate from the Bayesian procedure the median of the posterior PDF and the 16th and 84th percentiles as proxies for the range of $\pm 1\sigma$ uncertainties. The same procedure is then repeated for the (m$_{F606W}$-m$_{F814W}$, m$_{F814W}$) CMD (see Appendix~\ref{app:isofit}) and as our final values we take the mean of the best-fit parameters from the two cases, again following \citet{massari23,massari_25}.  This yields $\rm t=13.76^{+0.88}_{-1.30}$ Gyr,  $\rm (m-M)_0=16.85\pm 0.06$, $\rm E(B-V)=0.07\pm 0.01$ and $\rm [M/H]=-2.5\pm0.1$.

\section{Structural parameters} \label{sect:struc}

In this section we describe the methodology adopted to determine the structural parameters of Kop~2. Our analysis is based on the framework introduced by \citet{martin08,Martin2016}, which has been widely applied to sparse stellar systems \citep[see e.g.,][]{munoz18,Smith2023}. In Section~\ref{subsec:method} we introduce the model and the general formalism, in Section~\ref{subsec:data} we describe the subsample of stars used in the structural analysis, in Section~\ref{subsec:logl} we present the likelihood function adopted in the fit and Section~\ref{subsec:param} we discuss our results.

\subsection{The model}
\label{subsec:method}

%   intro
In our framework, the system is described by a probabilistic mixture model, $\PP$, composed of distinct components representing different physical populations and depending on a set of free parameters, $\btheta$. In our case, the model includes a term describing the cluster and a second term accounting for the foreground contamination, which we assume to be spatially constant. Given an observed dataset, which in our specific case consists of the projected positions of stars on the plane of the sky, the free parameters of the model are inferred by maximizing the likelihood of the model evaluated on the data. The model is defined as
\begin{equation}\label{for:prob}
    \PP(X,Y \mid \btheta) = \frac{w\,\Sigma(X,Y \mid \btheta)} {\int_\Area \Sigma(X,Y \mid \btheta)\dd X\dd Y}+\frac{(1-w)}{A},
\end{equation}
where $\Sigma$ represents the spatial distribution of stars in Kop~2, while the second term represents the foreground. The denominator of the first term, which we refer to as $\ASigma\equiv\int_\Area \Sigma(X,Y \,|\, \btheta)\dd X\dd Y$, ensures that the cluster density model is normalized to unity over the region $\Area$ covered by the observations. We have called $A$ the area of this region. We note that $\ASigma$ further represents the fraction of the Kop~2 model that is sampled by the spatial selection function and that it is directly used to infer the cluster total mass (see Section~\ref{sec:lf}). Finally, $0<w<1$ is the fractional contribution of the cluster density with respect to the total. 

%   the pseudoking
As cluster density, we adopt the flattened \citet[][K62]{king62} model
\begin{equation}\label{for:mod}
    \Sigma(m \mid \btheta) = \Sigmao\left[\frac{1}{\sqrt{1+\left(\frac{m}{\Rc}\right)^2}}-\frac{1}{\sqrt{1+\left(\frac{\Rt}{\Rc}\right)^2}}\right]^2,
\end{equation}
with $m$ the elliptical radius, defined as
\begin{equation}\begin{split}
    m^2 = & \left[\frac{(X-\Xo)\cos\phi - (Y-\Yo)\sin\phi}{1-e}\right]^2 + \\ &  \left[(X-\Xo)\sin\phi + (Y-\Yo)\cos\phi\right]^2. \\
\end{split}\end{equation}

Here, $(\Xo,\Yo)$ represents the offset of the cluster center with respect to a reference position. The parameter $e$ is the ellipticity, defined as $e=1-b/a$, where $a$ and $b$ are the semi-major and semi-minor axes, respectively. $\phi$ is the position angle (PA): $\phi=0$ corresponds to the major axis aligned with the $Y$-direction, and $\phi$ increases counterclockwise.

%   why a King.
A K62 model provides a flexible and convenient parametrization widely used to describe the spatial distribution of GCs. In its standard formulation, it represents a two-parameter family of models, defined by the core radius, $\Rc$, and the truncation radius, $\Rt$. In our flattened formulation, the core radius defines the elliptical central region where the surface density is approximately constant. The truncation radius sets the extent of the cluster by defining the outer boundary where the surface density drops to zero ($\Sigma(m>\Rt)=0$). It is often convenient to introduce an additional parameter, the concentration $c$, defined as $c=\logten(\Rt/\Rc)$, which provides a compact measure of the relative concentration of the system (K62).

%   comment on the total number of free parameter
Thus, the total number of free parameters of the model is seven, namely the offset centers, $(\Xo,\Yo)$, the core radius, $\Rc$, the concentration, $c$, the ellipticity $e$, the position angle, $\phi$, and the weight of the cluster contribution $w$.

\begin{figure*}[h]
    \centering
    \includegraphics[width=0.9\textwidth]{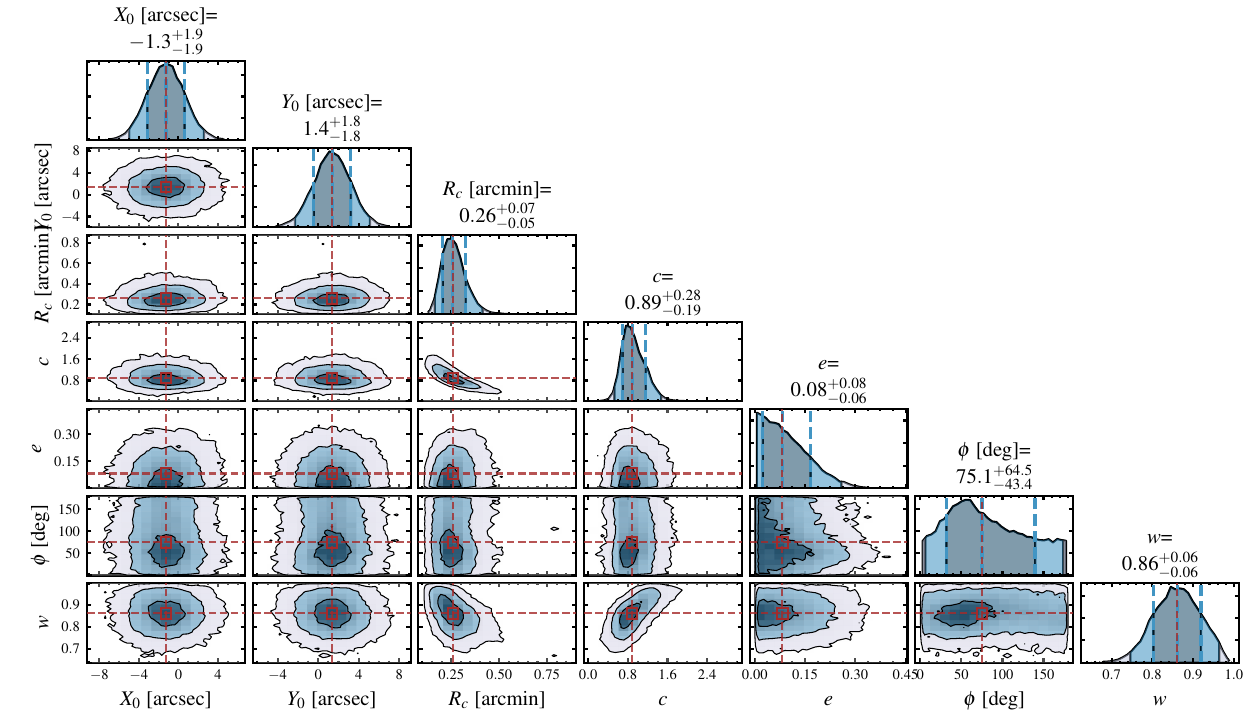}
    \caption{Marginalized one- and two-dimensional posterior distributions of the model free parameters. The contours in the two-dimensional marginalized distributions enclose, respectively, 68\%, 95\%, and 99\% of the total probability. In the one-dimensional marginalized distributions, the vertical blue lines indicate the 16th and 84th percentiles, while the vertical red line indicates the 50th percentile (median). The dashed red lines and squares in the two-dimensional marginalized distributions mark the position of the median model, i.e. the model whose parameters are given by the median values of the corresponding posterior distributions.}
    \label{fig:corner}
\end{figure*}

\subsection{The dataset}
\label{subsec:data}

We apply the method described above to a subsample of the available stars. The definition of this subsample is guided by a number of practical considerations: i) the definition of the integration region, i.e. the region $\Area$ in equation~(\ref{for:mod}) over which the model is constrained; ii) the removal of obvious non-member stars, to improve the accuracy of the cluster parameter inference; and iii) the definition of a magnitude completeness limit, to avoid introducing radial dependencies in the sampling of the stellar density. This latter condition ensures that, at any given magnitude, the selected stars trace the underlying density in a consistent way across the field.

%   points description
We first consider the definition of the integration region. As already noted in Section~\ref{subsec:method}, HST observations only provide a partial spatial coverage of the system, limited to the footprints of the ACS/WFC and WFC3/UVIS instruments. In our analysis, we further restrict this region by excluding small areas near the edges of both fields of view, as well as masking the interchip gaps with a width of 8 arcseconds in both instruments. These regions are removed to mitigate strong completeness variations arising from non-uniform coverage at the field edges and from the dithering pattern across the detectors.

%   removal of non members
We then restrict the sample by removing likely non-member stars. As shown in Figs~\ref{fig:CMD} and \ref{fig:agedist}, the cluster sequence is well separated from the foreground contamination, allowing a substantial fraction of non-members to be excluded with high confidence. We therefore define a broad selection region around the best-fit isochrone of Fig.~\ref{fig:agedist} and retain only the stars falling within this region, thereby increasing the contrast between the cluster and background densities and improving the precision of parameter inference \citep{Munoz2012}. This step is not intended to fully remove the contamination, but rather to mitigate its impact; accordingly, the model retains both cluster and foreground components. Then we apply all the quality cuts that were adopted to study the completeness and to derive the LF, described in Appendix~\ref{app:cf}. Finally, we apply a magnitude cut to remove stars with ${\rm m}_{\rm F606W}>25.0$. This choice is motivated by completeness considerations, based on the analysis of artificial star tests; we refer to Appendix~\ref{app:cf} for a detailed description of the artificial star catalogue and the criteria adopted to define a reliable magnitude completeness limit. 
At the end of the procedure, the dataset, $\DD$, consists of $N$ stars with measured position $(\Xj,\Yj)$, with $i=1,...,N$.

\begin{figure*}
    \centering
    \includegraphics[width=0.9\textwidth]{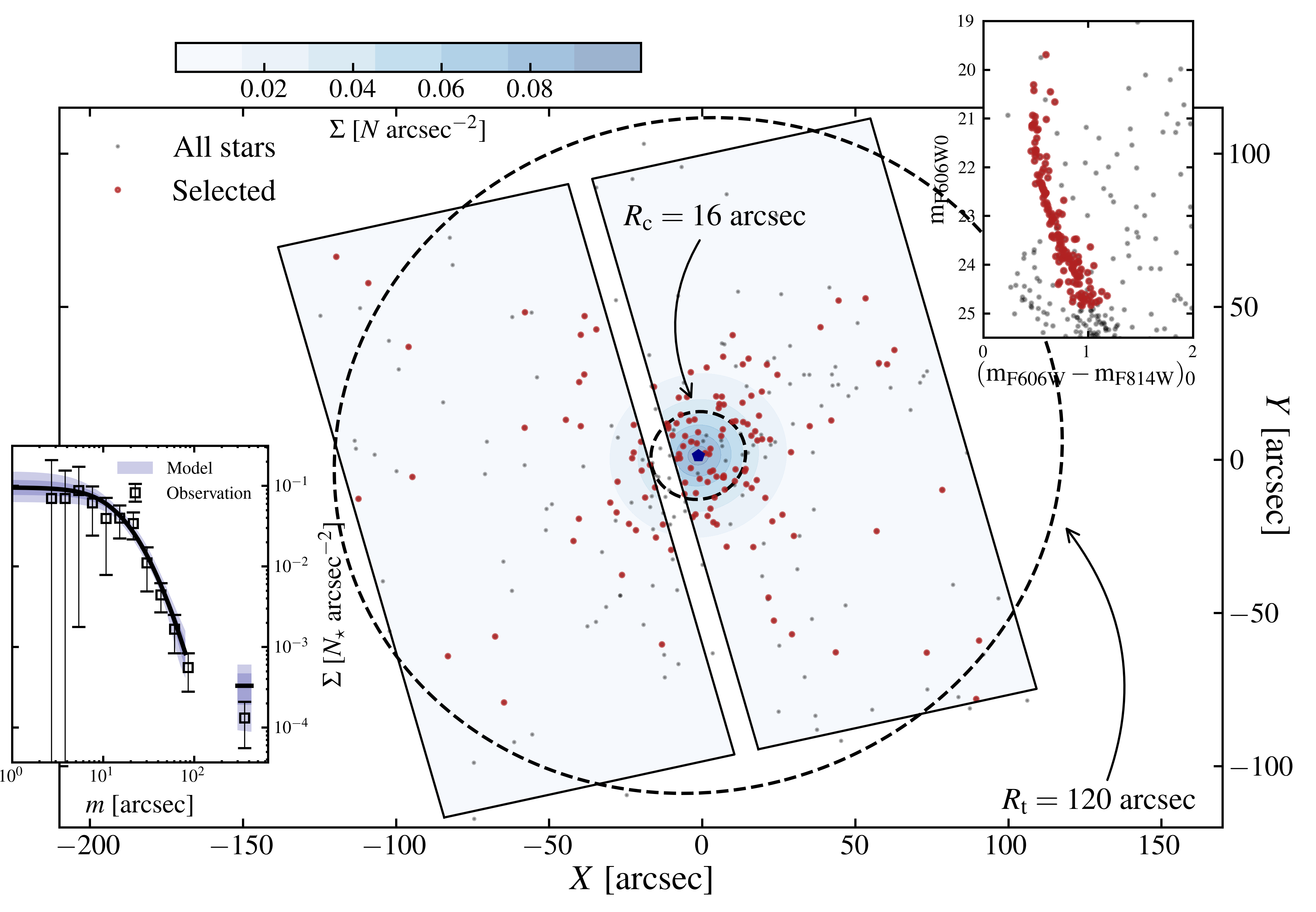}
    \caption{Main panel: spatial distribution of all stars passing the quality cuts (black) and of the subsample used for the structural fit (red). The two rectangles mark the ACS subregions included in the analysis; the WFC3 field is not shown for clarity. The underlying blue-scale map illustrates the density variations of the median model, with darker shades corresponding to higher densities; the darkest pentagon marks the inferred center. The dashed ellipses denote the core and truncation radii, as derived from the median model. The median model is defined as the model whose parameters are given by the median of the corresponding one-dimensional posterior distributions, and it is shown in Fig.~\ref{fig:corner} with a red square. Top-right panel: color–magnitude diagram of the full sample (black), and of the stars above the cut used in the reference fit (red). Bottom-right panel: median density profile (along the major axis), with $1\sigma$ and $2\sigma$ uncertainties inferred from the fit. The model profile is compared to the observed one, computed in elliptical bins with flattening and orientation fixed to the reference model.}
    \label{fig:spatial}
\end{figure*}

\subsection{The likelihood}
\label{subsec:logl}

%   the likelihood
The probabilistic model introduced in equation~\ref{for:prob} naturally defines a likelihood function once a dataset is specified and interpreted as a realization of the model. The likelihood of the model, given the dataset, is 
\begin{equation}\label{for:likl}
    \LL(\btheta\mid\DD) = \prod_{i=1}^{N}\PP(\Xj,\Yj\mid\btheta)
\end{equation}
i.e, the product of the individual probabilities evaluated at the position of each star. The posterior probability distribution of the model parameters, $p$, is obtained through Bayes’ theorem and is 
\begin{equation}
    p\,(\btheta \mid \DD) \propto \LL(\DD \mid \btheta)\, \Pi(\btheta),    
\end{equation}
i.e., proportional to the product of the likelihood and the prior, $\Pi$. The exploration of $p$ is carried out using a MCMC technique, implemented with the affine-invariant ensemble sampler {\tt emcee}. In our analysis, we adopt 56 walkers and generate chains of length 6000 steps. To ensure convergence, an initial burn-in phase of 500 steps is discarded. We further remove walkers that do not reach convergence, assessed through the behavior of the log-posterior. The remaining chains are then thinned by a factor of 10, corresponding to approximately half of the median autocorrelation time, in order to reduce short-range correlations and limit the computational cost associated with the evaluation of derived quantities. From the cleaned posterior distributions, we adopt the median of each parameter as the reference value, while credible intervals are computed from the 16th and 84th percentiles (1$\sigma$) and from the 5th and 95th percentiles (2$\sigma$). 

\subsection{Newly derived structural parameters.}
\label{subsec:param}

%   coner
Fig.~\ref{fig:corner} shows the marginalized one- and two-dimensional posterior distributions obtained from the structural fit, illustrating the constraints on the model parameters and their correlations. The inferred parameter estimates, together with the corresponding derived quantities, are reported in Table~\ref{tab:param}.

%   comment on center
We derive a new estimate of the centre of Kop 2, finding RA=$119.5705^\circ$ and Dec=$26.2558^\circ$, with corresponding uncertainties of the order of $1.8$ arcsec in both coordinates (see Table~\ref{tab:param}). Our determination is significantly more precise than previous literature estimates: compared to \cite{munoz18}, the uncertainties on both coordinates are reduced by approximately a factor of two, while the two measurements remain fully consistent within $1\sigma$. We also find excellent agreement with the centre reported by \cite{koposov_07} and \cite{paust14}, who, however, did not provide uncertainties on their estimate.

%   other params
Our estimates for the K62 core radius and concentrations are $\Rc=0.26_{-0.05}^{+0.07}$ arcmin and $c=0.89_{-0.19}^{+0.28}$, respectively, that translate into a truncation radius of $\Rt=1.99_{-0.43}^{+1.11}$ arcmin and a half-light radius of $\Rh=0.39^{+0.06}_{-0.04}$ arcmin. The system appears remarkably circular, with a measured ellipticity of $e=0.08^{+0.08}_{-0.06}$. However, the posterior distribution is strongly peaked toward the lower prior boundary at $\epsilon=0$ (see Fig.~\ref{fig:corner}), indicating that the data only place an upper limit on the flattening rather than supporting a significant detection of flattening.

%   side note on use of different models
To test the robustness of the structural parameters inferece, we repeated the analysis using alternative density profiles, namely a flattened \cite{Plummer1911} and exponential models. In the case of Kop 2, all models return fully consistent estimates of the centre, half-light radius, and flattening, with statistically equivalent uncertainties. A broader comparison among different profile families, extended to additional MGCS clusters, will be presented in a forthcoming paper.

%   comment to Fig 5
The main panel of Fig.~\ref{fig:spatial} shows the spatial distribution of all bona-fide, well-measured stars in the ACS/WFC footprint together with the subset of sources used in the fit, namely those passing the quality selections described in Section~\ref{subsec:data}. For clarity, we do not show the parallel WFC3/UVIS field, which end up containing only three stars after applying the mentioned selection criteria. The top right inset shows the subset of stars in the CMD, while the bottom left inset shows a comparison between the observed radial density profile with the median K62 model and its $1\sigma$ confidence interval. The agreement between model and data is excellent over the full radial range probed by the observations, and it confirms that the adopted profile provides a reliable description of the system. The stellar distribution also appears largely circular, in agreement with the low inferred ellipticity. The inferred truncation radius falls between the ACS/WFC footprint and the WFC3 parallel field, indicating that our observations sample the cluster out to its outskirts and therefore provide solid constraints on its structural extent.

%   comparison
A comparison with previous structural studies is necessarily limited, since, to our knowledge, \citet{munoz18} provides the only earlier systematic analysis that explicitly accounts for the possible flattening of the system. Overall, we find good agreement with the structural parameters inferred by the authors. The best-fit K62 values of $\Rc$ and $\Rt$ reported by \citet{munoz18} are fully consistent with our measurements within $1\sigma$, although our estimates are significantly more precise.\footnote{Although the two determinations of $\Rt$ are statistically consistent within $1\sigma$, our median truncation radius is smaller than that reported by \cite{munoz18}. We note that the quoted uncertainties are not directly comparable, since our errors are derived from the Bayesian posterior distribution, whereas \cite{munoz18} estimated them through bootstrap resampling. Nevertheless, our analysis generally provides tighter constraints on the structural parameters.} The main discrepancy instead concerns the inferred flattening of the system. While our analysis favours a nearly circular configuration, \cite{munoz18} reported a significantly larger ellipticity ($e\simeq0.41$), incompatible with our estimate within the quoted uncertainties. This difference may plausibly arise from the different { depth, completeness and field of view} of the two data sets, which can affect the outer spatial distribution of stars and therefore bias the inferred shape of the system. 

%   effect of distance
Our revised distance estimate directly affects the inferred physical size of Kop 2. Adopting $D=23.4\pm0.6$ kpc, the physical half-light radius is $\Rh = 2.7^{+0.4}_{-0.3}$ pc, substantially smaller than that provided by, e.g., \cite{munoz18} of $\Rh=4.4\pm0.7$ pc, based on a significantly larger distance estimate. 

%--------------------
\begin{figure}[h]
\centering
\includegraphics[width=\columnwidth]{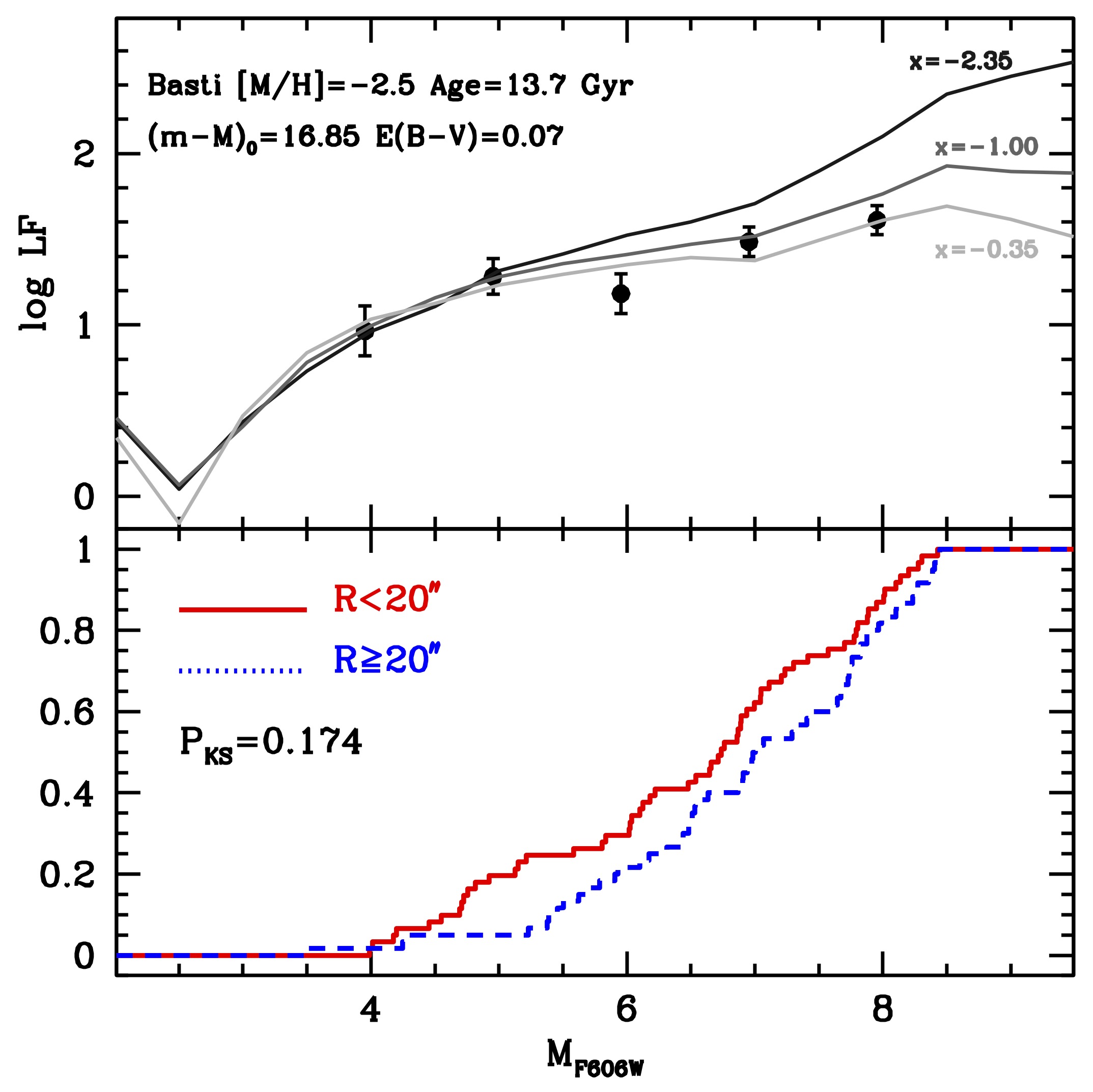}
\caption{Upper panel: the completeness-corrected LF of Kop~2 is compared with theoretical models with the same metallicity and age as the best-fitting isochrone and different slopes $x$ of the underlying IMF $\Phi(m)\propto m^{x}$. The theoretical LFs are normalised to the mean of the two brightest points of the observed LF. Lower panel: cumulative observed LFs of the subsample of Kop~2 stars lying within (red continuous line) and outside (blue dashed line) a circle of radius $R=20.0\arcsec$ centred on the centre of the cluster ($R\le 60.0\arcsec$). The two subsamples contains 61 and 60 stars, respectively.  \label{fig:lf}}
\end{figure}
%--------------------

\section{Luminosity function, integrated luminosity, mass segregation}
\label{sec:lf}

In the upper panel of Fig.~\ref{fig:lf} we compare the observed LF, corrected for completeness according to the Completeness Fraction (CF) described in Appendix~\ref{app:cf} with a set of theoretical BaSTI models with the same age and metallicity of the best-fit isochrone of Fig.~\ref{fig:agedist}, and adopting the best-fit distance modulus and reddening. The LF has been obtained from stars selected to lie around the best-fit isochrone of Fig.~\ref{fig:agedist}, to satisfy the quality cuts described in Appendix~\ref{app:cf} and limited to $\rm m_{F606W}<25.5$ to remain in the range where $CF\ge 0.5$ and there is no sign of radial variation of the completeness (see Appendix~\ref{app:cf}). According to the best-fit isochrone of Fig.~\ref{fig:agedist}, our LF includes stars in the range of mass from $0.72~M_{\sun}$, at $\rm m_{F606W}=20.5$, to $0.38~M_{\sun}$, at $\rm m_{F606W}=25.5$.

The observed LF is reasonably reproduced by the theoretical LF  of a populations with mass function, in the form $\frac{dN}{dm_{\star}}\propto m_{\star}^{x}$, with $x\ge-1.0$ \citep[where $x=-2.35$ corresponds to the][MF]{salp55}. The model with $x=-0.35$ provides the best overall representation of the observed LF and we take this as our reference model. \citet{paust14} found that their LF was best fitted by a Salpeter MF but this conclusion was based on shallower ground-based data sampling a much narrower range of mass compared to the HST photometry used here ($0.63~M_{\sun}\le M_{\star}\le 0.89~M_{\sun}$ vs. our LF sampling $0.38~M_{\sun}\le M_{\star}\le 0.73~M_{\sun}$). Moreover, they compared the observed LF with theoretical models with a widely erroneous metallicity, [Fe/H]=-0.6 and [$\alpha$/Fe]=+0.2, and a distance modulus $\rm (m-M)_0=17.75$, 0.9 mag larger than the one measured here.

$x=-0.35$ is among the flattest MFs inferred in Galactic GCs, generally interpreted as a sign of significant preferential depletion of low mass stars by the combined effect of mass segregation by two-body relaxation and severe pruning by Galactic tides \citep[see][and references therein]{paust09, paust10A, sollima17}. The low concentration and surface brightness of Kop~2 (Sect.~\ref{sect:struc}) are typical of globular clusters displaying similarly flat MFs \citep{paust10A}. On the other hand, significantly steeper stellar mass functions are typically inferred in UFDs  \citep[see][and references therein]{imf26}.

The normalization of the $x=-0.35$ model provides a direct estimate of the luminosity sampled by our data, in the sound hypothesis that the model LF is a proper description of the stellar population of Kop~2. We verified that the adoption of a distance at $\pm 1\sigma$ from our best-fit value has a negligible effect on the normalisation. We found that our dataset samples $\ASigma = 0.8695^{+0.013} _{-0.018}$ of the total light of the K62 model that best-fits the spatial distribution of our stars (see Sect.~\ref{sect:struc}), hence, by coupling this fraction with the sampled luminosity we got an absolute integrated magnitude of $\rm M_{F606W}=-1.06 \pm 0.22$, that we converted into
$\rm M_{V}=-0.95 \pm 0.22$ with the transformation by \citet{galleti06} . This is in excellent agreement with the independent estimate by \citet[][$\rm M_{V}=-0.9 \pm 0.8$, with a similar method on shallower ground-based data]{munoz18}, but with a significantly reduced uncertainty, owing to the high quality of the HST data used here. Assuming the mean stellar mass-to-light ratio of Galactic globular clusters from \citet{baumgardt_20}, $\frac{M_\star}{L_V}=1.83\pm 0.03$\footnote{ Please note that this includes the contribution of dark remnants, assuming that their contribution to the stellar mass is similar to Galactic GCs, since the $M/L$ ratio from \citet{baumgardt_20} are derived from dynamical estimates of the cluster mass. }, and M$_{V,\sun}=4.82$\footnote{From \url{https://mips.as.arizona.edu/~cnaw/sun.html}}, 
from our new measure of $M_V$ we obtain a total stellar mass of $M_{\star}=372\pm42~M_{\sun}$.  The mean mass of individual stars in the $x=-0.35$ model shown in Fig.~\ref{fig:lf} is 0.40~$\rm M_{\sun}$. Given this value, in the hypothesis of a pure stellar system (i.e., without dark matter) the two-body relaxation time of Kop~2, computed as in C26, is $t_{\rm  relax}=0.19$~Gyr, much shorter than the age of the system. 

Under the same hypothesis, the updated stellar mass and distance allow us to revise the expected Jacobi radius of the cluster. For a stellar system of mass $M_\star$ orbiting within the Galactic potential, the Jacobi radius can be written as \citep{king62}
\begin{equation}\label{for:jac}
    \RJ\equiv\biggl(\frac{GM_\star}{2V_{\rm G}^2}\biggr)^{1/3}\,R_{\rm GC}^{2/3},
\end{equation}
where $R_{\rm GC}$ denotes the galactocentric distance of the system, $V_{\rm G}$ is the MW circular velocity at that radius, and $G$ is the gravitational constant. The Jacobi radius corresponds to the characteristic tidal boundary imposed by the Galactic potential and thus provides a useful indicator of the dynamical influence of the MW tidal field on the system. By applying equation~(\ref{for:jac}) using the mass and galactocentric distance inferred in this work, and adopting the Galactic potential model of \cite{mcmillan_17} to compute the circular velocity, we derive a Jacobi radius of $\RJ = 25.2^{+1.0}_{-1.0}$ pc. Using the intrinsic three-dimensional half-mass radius, which for K62 models is approximately related to the projected half-light radius through $\rh \simeq 1.33\Rh$, we obtain a ratio $\rh/\RJ = 0.14_{-0.01}^{+0.02}$. According to the classification of \cite{Baumgardt2010b}, this would place the system at the lower boundary of the tidally filling cluster regime, i.e. systems that are expected to be affected by the Galactic tidal field and may be close to dissolution.

In the lower panel of Fig.~\ref{fig:lf} the cumulative LF of the stars within (red continuous line) and outside (blue dashed line) $R=20.0$ arcsec from the centre of the cluster are compared. The radial threshold has been chosen to have approximately the same number of stars in the two subsamples and the outer sample is limited to $R\le 60.0$ arcsec to minimise possible effects of contamination from non-member stars.
The $\rm M_{F606W}$ distribution is more skewed toward brighter (more massive) stars in the inner region of the cluster than in the outer part, suggesting that some degree of mass segregation may be present.
However, a Kolmogorov-Smirnov test do not detect any statistically significant difference between the two distributions, likely due to the low number of involved stars. { The lack of a statistically significant signal of mass segregation is confirmed by the ratio of the median distance from the centre of a sample of bright and faint stars $\rm r=R_{H,bright}/R_{H,faint}$, as defined in \citet{baum22}. We computed r on the stars used in Fig.~\ref{fig:lf} with $\rm m_{F606W}\le 25.0$ and we adopted the threshold at $\rm m_{F606W}= 23.6$ to have bright and faint subsamples of the same size. Following \citet{baum22}, we obtain $r=1.02\pm 0.19$, in agreement, within the uncertainties, with the value obtained by the same authors from a different, ground-based sample ($r=0.73\pm 0.17$; see also Appendix~\ref{app:evol}, for comparison with models).}

%-----------------
\begin{table}[h]
\caption{Measured parameters of Koposov~2.}
\label{tab:param}
\centering
\renewcommand{\arraystretch}{1.35}
\footnotesize
\begin{tabular}{lrr}
\toprule
\toprule
Parameter & Value &  Prior  \\
\midrule
\midrule
Age [Gyr] & $13.76^{+0.88}_{-1.30}$ & $\UU(0, 15)$ \\
Age [Gyr] (P$>$5\%)$^a$ & $>11.6$ & $\UU(0, 15)$ \\
E(B-V)    & $0.07\pm 0.01$       & $\NN(0.05,0.1)$ \\ 
${\rm [M/H]}$  & $-2.5\pm 0.1$ & $\NN(-2.5,0.1)$ \\
$(m-M)_0$ & $16.85\pm 0.06$ & $\NN(17.6,0.5)$ \\
$D$ [kpc]   & $23.4\pm0.6$ & --  \\
\midrule
RA [deg] &  $119.5705_{-0.0005}^{+0.0005}$ & $\UU(-50, 50)$ \\
Dec [deg] & $26.2558_{-0.0005}^{+0.0005}$ & $\UU(-50, 50)$  \\
$\Rc$ [arcmin] & $0.26_{-0.05}^{+0.07}$ & $\UU(0, 2)$ \\
$\Rh$ [arcmin] & $0.39_{-0.04}^{+0.06}$ & - \\ 
$\Rt$ [arcmin] & $1.99_{-0.43}^{+1.11}$ & - \\
$c$ & $0.89_{-0.19}^{+0.28}$ & $\UU(0, 3)$ \\
$e$ & $0.08^{+0.08}_{-0.06}$ & $\UU(0, 1)$ \\
$e$  (P$<$95\%)$^a$ & $<0.23$ & $\UU(0, 1)$ \\
$\phi$ [deg]  & $75.1_{-43.4}^{+64.5}$ & $\UU(0, 180)$ \\
$w$  & $0.86_{-0.06}^{+0.06}$ & $\UU(0, 1)$ \\
\midrule
$\Rc$ [pc] & $1.8_{-0.4}^{+0.5}$  & - \\
$\Rh$ [pc] &  $2.7_{-0.3}^{+0.4}$ & - \\
$\Rt$ [pc] & $13.6_{-3.0}^{+7.6}$ & - \\
M$_V$ & $-0.95\pm 0.22$ & - \\
$M_{\star}^{b}$ [$M_{\sun}$] & $372\pm 42$ & - \\
\bottomrule
\bottomrule
\end{tabular}
\tablefoot{In the case of uniform priors, $\UU(a,b)$ denotes a uniform distribution between the lower and upper bounds $a$ and $b$, while $\NN(\mu,\sigma)$ denotes a Gaussian distribution with mean $\mu$ and standard deviation $\sigma$. Quantities without a prior have been derived from the models parameters. 
{ $^a$ For the two parameters for which the mode of the posterior PDF is reached at one limit of the prior PDF we also report the 5th (Age) and 95th ($e$) percentiles of the one-dimensional posterior distributions, that should be interpreted as one-sided lower or upper limits. }
{}$^{(b)}$ Obtained from $M_V$ by adopting the mean mass-to-light ratio of Galactic globular clusters, $M/L_V=1.83\pm0.03$, from \citet{baumgardt_20}. }
\vspace{2pt}
\end{table}
%-----------------

%--------------------
\begin{figure}[h]
\centering
\includegraphics[width=1.0\columnwidth]{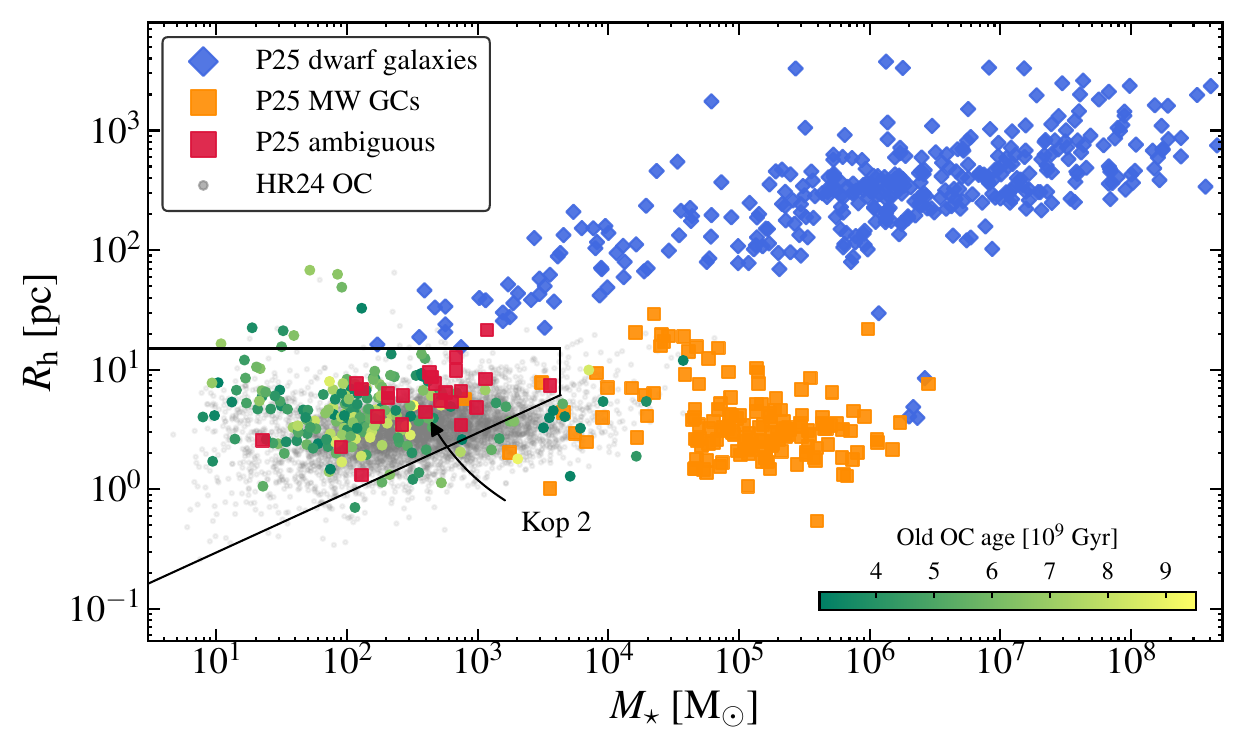}
\caption{Low mass local stellar systems in the plan opposing the stellar mass  to the half-light radius. Blue diamonds are dwarf galaxies in the Local Volume form P25.
%\citet[][P25 hereafter]{pace_25} 
Orange squares are Galactic globular clusters from the lists of "Harris GC" and "New MW GCs" by P25, while red squares are from the list of "ambiguous" systems by P25. 
The parameters of Kop~2 have been updated with the measures presented in this paper. The cluster is indicated by an arrow and labelled.
Grey dots are all the Open Clusters in the catalogue by \citet[][HR24]{hr24}. We used the radius containing 50\% of the cluster members $r_{50,J}$ from HR24 as a proxy for the half-light radius for OCs. The HR24 OCs with age$> 3.0$~Gyr are highlighted as small filled circles colour-coded according to their age. The thin lines enclose the portion of the diagram where UFCS are located according to the definition by C26. }
\label{fig:mrh}
\end{figure}
%--------------------

\section{The nature of Kop~2} 
\label{sect:disc}

Following the classification criteria proposed by C26, a very old age is not a discriminant between dwarf galaxies and GCs, while the very low metallicity would support the galaxy hypothesis. In particular, Kop~2 fits well into the M$_V$-[Fe/H] relation of local dwarf galaxies, at its faint end, close to the UFDs Tucana~V and Draco~II \citep{pace_25}. However, as noted before, this criterion is circumstantial, as there is evidence that GCs as metal-poor as UFDs likely existed in the Galactic halo \citep[see, e.g.,][their Fig.~25, in particular]{iba24}.

The new observational evidence provided in the present study that is most promising for ascertaining the nature of Kop~2, is the inference of a pretty flat ($x>-1.0$) present-day mass function,
suggesting a strong selective depletion of faint stars, typically expected \citep{vh97,bm03} and observed to occur in some star clusters that lost a significant fraction of their initial mass due to the combined effects of two-body relaxation and the Galactic tidal field \citep{paust09,paust10A,sollima17}.
Such a flat present-day MF lends support to a classification of Kop 2 as a dissolving GCs since UFDs are expected to be less affected by the dynamical processes flattening the IMF and, indeed, typically show steeper IMFs, with $x \lesssim -1.0$ and reaching $x \lesssim -2.0$, as recently reviewed by \citet{imf26}.

As discussed in  Section 5, our estimate for the filling factor $(\rh/\RJ \sim 0.14)$ for Kop 2 puts this system in the group of tidally filling systems \citep{Baumgardt2010b} that may be approaching dissolution. This is also consistent with the flat present-day MF, another possible fingerprint of a significant preferential loss of low-mass stars. In Appendix~\ref{app:evol} we show the results of a suite of $N$-body simulations following the dynamical evolution of a small GC exposed to a weak tidal field. While a much larger (and tuned) survey of simulations would be necessary to provide a specific fit to the properties of Kop 2 to identify the possible range of viable initial conditions, the models presented in Appendix D serve to illustrate the possible evolutionary path of a GC evolving towards dissolution over $\simeq 14$~Gyr into a present-day dwarf-depleted very low-mass system at large galactocentric distances, without requiring the additional binding support from a dark matter halo.

Finally, in Fig.~\ref{fig:mrh} we plotted the dwarf galaxies in the Local Volume, the Galactic GCs and the stellar systems classified as ambiguous from the recent \citet[][P25 hereafter]{pace_25}  dataset { (version v1.0.1)}. It is shown that, according to the newly derived structural parameters, Kop~2 is
significantly more compact than { confirmed} UFDs of similar stellar mass. To give a direct taste of the difference, the two UFDs being most similar to Kop~2 in terms of stellar mass mentioned above, Draco~II and Tucana~V, have $\Rh=18.7_{-3.7}^{+2.8}$~pc and $\Rh=33.0_{-8.1}^{+8.9}$~pc, respectively (P25), to be compared with $\Rh=2.7^{+0.4}_{-0.3}$~pc of Kop~2. { While compactness, like, e.g., metallicity is a circumstantial criterion to discriminate clusters from galaxies, the small $\Rh$ of Kop~2} suggests that the system is more akin to star clusters than to dwarf galaxies.

\subsection{The origin of Kop~2}
\label{subsec:origin}

The sector of Fig.~\ref{fig:mrh} where UFCS resides, according to the definition by C26, is the lower left corner enclosed by thin black lines. 
It is very interesting to note that the UFCS portion of the diagram encloses not only virtually all the P25 ``ambiguous'' systems and some faint system classified as GC, but also the vast majority of Open Clusters (OC)\footnote{The inclusion of OCs in this kind of comparisons was introduced by \citet{torrealba_19} and \citet{mau_19}, but in the $M_V$ - $\Rh$ diagram instead of the $M_{\star}$ - $\Rh$ diagram shown here. $M_V$ is a good proxy for stellar mass when comparing stellar systems of similar age but it may be misleading
in presence of significant age differences, like those, e.g., between most OCs and some UFCS, due to the age dependency of the mass-to-light ratio.}. OC data were taken from the catalogue by \citet[][HR24]{hr24}\footnote{ It may be worth noting that all the OCs in the HR 24 catalogue have $\rm |Z|<1.0$~kpc, while UFCS are halo objects, by definition ( $\rm |Z|>5.0$~kpc, C26)}, using the radius containing 50\% of the cluster members ($r_{50,J}$) as a proxy for $\Rh$. The long-lived OCs, those with age from 3.0~Gyr to $\simeq 9.5$~Gyr, overlap very well with the UFCS having $M_{\star}\lesssim 10^3~M_{\sun}$. 
Since it is clear that the UFCS class includes objects of various nature (e.g., UFDs or GCs, C26), the fact that OCs of any age share the structural properties of the class suggests a possible additional evolutionary channel. Some open clusters may have been ejected from the MW disc in a wide high-$Z_{\rm max}$ orbit, by, e.g., the impact of an existing or a disrupted satellite with the Galactic Disc \citep[like Sgr~dSph or Gaia-Sausage-Enceladus, see, e.g., ][and references therein]{helmi_18,ceccarelli_25}. In the new off-disc orbit the ejected cluster could have been able to survive until the present day, unlike OCs permanently residing in the disc. If this channel is actually viable for a given UFCS should be investigated with proper dynamical modelling, given the orbit.

However, in spite of its high $L_Z$ orbit, this does not seem a likely formation channel for Kop~2. First of all, given its very low metallicity, it should have been ejected at an early epoch from a very pristine metal-poor disc, whose existence in the early Milky Way is far from certain \citep[but see, e.g.,][and references therein]{mic24,xiang25,laporte26,highzdisc26}. Second, and more relevant, it seems unlikely that a stellar system can be brought on such a wide orbit from an early disc whose radial size should not have much exceeded $\sim 10$~kpc \citep{kawadisc26,xiang25}. 

On the other hand, the wide, high $Z_{\rm max}$ orbit and the very low metallicity strongly suggest that if Kop~2 is a star cluster it was likely brought into the outer MW halo by a former and now disrupted dwarf galaxy satellite. In a recent analysis of the NEFERTITI set of detailed formation models of MW analogues, \citet{kotsou26} found that the fraction of accreted stars increases with decreasing metallicity, in good agreement with previous studies. In particular, for $\rm [Fe/H]<-3.0$ nearly all the stars presently bound to their MW analogues were accreted from previous dwarf satellites. According to these models, it is highly improbable that a cluster with $\rm [Fe/H]=-2.9$ like Kop~2 was formed in-situ. \citet{ceccarelli_26} did not find any obvious association of Kop~2 with known relics of accretion events (see also C26), but it must be recalled that most of those have been detected in the realm where Gaia DR3 \citep{gaia_23} astrometry is most accurate, that is within $R_{\rm GC}\lesssim 10$~kpc, while the entire orbit of Kop~2 ranges beyond $R_{\rm GC}\gtrsim 28$~kpc up to $R_{\rm GC}\gtrsim 60$~kpc (Appendix~\ref{app:orb}). The dwarf galaxy that possibly brought in Kop~2 may have been quite faint and/or disrupted long ago. 
On the other hand, the likely association of the UFCS Koposov~1 with the disrupting Sgr dSph \citep[][C26]{ceccarelli_26}, provides a direct proof that such faint clusters can originate in dwarf galaxies and be long-lived in spite of their small mass (note that Koposov~1 has been classified as star cluster by C26).

\section{Summary and conclusions}
\label{sec:conclu}

We used deep HST photometry from MGCS to significantly improve our knowledge of the faint stellar system Koposov~2. We obtained robust measures of the distance, age, size and absolute integrated magnitude of the system, much more precise and accurate than those previously available. If the star for which a spectroscopic measure of the metallicity was obtained by \citet{geha_26}, C26 and \citet{ceccarelli_26} is indeed a real member of Koposov~2, which appears as extremely likely, the system is very metal-poor ([Fe/H]=-2.9) and extremely old (age $\gtrsim 13$~Gyr). The results of the measurements performed in this study are summarised in Table~\ref{tab:param}.

As a summary of our discussion, the new observational evidence presented here supports the idea that Kop~2 is a genuine star cluster, but are not yet sufficient to ultimately exclude the possibility that it can be an extremely faint and compact UFD \citep[that likely lost much of its dark matter halo,][]{errani24,ss26}. The final tackling of this problem for these most ambiguous cases, like Kop~2, still resides in very challenging spectroscopic observations to obtain accurate and reliable measures of the velocity and metallicity dispersions. 

The former appears as the most difficult to achieve since these faint and distant systems have typical $ \sigma_{V_{\rm los}}\lesssim 3.0-5.0~\rm km~s^{-1}$, with predicted values of $\sigma_{V_{\rm los}}\lesssim 1.0~\rm km~s^{-1}$ in case of pure stellar content (C26). Obtaining reliable measures of such low velocity dispersions would require medium-high resolution spectroscopy of as many member stars as possible (where reasonable numbers are of the order $<10-20$ within a circular region of radius $\lesssim 1-2$ arcmin). { However, in our view, it would be also very important to obtain} several repeated observations of the same stars both to evaluate the impact of binary systems in inflating $\sigma_{V_{\rm los}}$ and to get accurate empirical evaluation of the uncertainty on the individual velocity of single stars. This is crucial for the unbiased measure of the intrinsic dispersion by the state-of-the-art Bayesian inference methods used today (e.g., C26, and references therein), in particular for very low $\rm \sigma_{V_{\rm los}}$ systems like Kop~2. Moreover, good measures may still be insufficient to provide a final clear-cut answer, as demonstrated by the controversial case of UMa~III/UNIONS~1 \citep{smith24,errani24,devlin25,rostami25,arroyo26,cerny_u26}.

On the other hand, $\sigma_{\rm [Fe/H]}$, while being nearly as observationally challenging as $\sigma_{V_{\rm los}}$, may be a more faithful diagnostic, since it keeps record of the past history of the considered system. The presence of a significant metallicity spread would indicate that in the past the stellar system was sufficiently massive to retain the ejecta of supernovae, making the diagnostic effective also for disrupting or disrupted stellar systems (see, e.g., 
\citealt{mic12,ws12,iba24} and references therein, but see also \citealt{Peng2006}).

Regardless of the actual nature of Kop~2, it is important to note that it may be the most metal-poor Galactic GC that is still bound or one of the most metal-poor UDFs, since the record holder, according to \citet{geha_26}, is 
Pictor~II which has $\rm \langle [Fe/H]\rangle =-2.99\pm 0.06$ \citep{pacepic25}, indistinguishable within the uncertainty from $\rm [Fe/H]=-2.91\pm 0.12$ of Kop~2. Eridanus~III is classified as ambiguous by \citet{simon24}, who find $\rm [Fe/H]=-3.10\pm 0.37$ and as ``candidate galaxy'' by C26, who find $\rm \langle [Fe/H]\rangle=-3.32\pm 0.21$.

\begin{acknowledgements}
This work has made use of data from the HST Treasury programme entitled {\em The Hubble - Missing Globular Cluster Survey} (GO-17435, PI: D. Massari). Data products are available at \url{https://www.oas.inaf.it/en/research/m2-en/mgcs-en/}
Based on observations with the NASA/ESA HST, obtained at the Space Telescope Science Institute, which is operated by AURA, Inc., under NASA contract NAS 5-26555. Support for Program number GO-17435 was provided through grants from STScI under NASA contract NAS5-26555. This work has made use of the Local Volume Database\footnote{\url{https://github.com/apace7/local_volume_database}} \citep{pace_25}.
\\

We are grateful to an anonymous Referee for a constructive report that helped us to improve the quality of the paper.

\\
RP acknowledges the support to this study by the INAF Mini Grant 2025 (Ob.Fu.1.05.24.07.05, CUP C33C24001390005). RP also acknowledges funding from the INAF Theory 2024 Grant project: Magnetohydrodynamic Simulations of Galactic Molecular Clouds: Resolving Stellar Birth and Proto-planetary Discs with an Enhanced Chemical Network

MDL acknowledges financial support from the project {\em LEGO – Reconstructing the building blocks of the Galaxy by chemical tagging} (PI: Mucciarelli) granted by the Italian MUR through contract PRIN2022LLP8TK\_001.

A.B. acknowledges support from STScI grant GO-17435.

SS acknowledges funding from the European Union under the grant ERC-2022-AdG, ‘StarDance: the non-canonical evolution of stars in clusters’, Grant
Agreement 101093572, PI: E. Pancino.
\end{acknowledgements}
%--------------------------------------------------------------------
\bibliographystyle{aa} % style aa.bst
\bibliography{biblio} % your references Yourfile.bib

\begin{appendix}

\section{Completeness}
\label{app:cf}

For all the applications for which we made use of star counts, like, e.g., the modelling of the surface density distribution (Sect.~\ref{sect:struc}) or the LF (Sect.~\ref{sec:lf}), we adopted a simple  set of criteria to select bona-fide well-measured stars that can be easily reproduced when deriving the completeness fraction (CF) from artificial star test \citep[][and references therein]{mic02}.

In particular we retained in our final catalogues only stars having:

\begin{itemize}
    \item{} $\sigma_{\rm mag}<0.25$ in both $\rm m_{F606W}$ and $\rm m_{F814W}$ magnitudes

    \item{} {\tt QFIT$_{\rm F606W}<0.5$} and {\tt QFIT$_{\rm F814W}<0.5$}

    \item{}  {\tt |RADXS|}$_{\rm F606W}<0.25$ and {\tt |RADXS|}$_{\rm F814W}<0.25$

    \item{} $N_u/N_f\ge 0.5$ in both F606W and F814W

    \item{} $o_{\rm F606W}<1.0$ and $o_{\rm F814W}<1.0$
    
\end{itemize}

where the meaning of the various parameters is defined in \citet{libra26}.

Concerning artificial stars, we consider an artificial star as successfully recovered if the input and output positions match, if it  satisfies all the conditions above and if

\begin{itemize}
    \item {} $mag_{out}-mag_{in}>-0.75$ from both magnitudes
\end{itemize}

that is, an artificial star of magnitude $m$ falling over a real star brighter than $m$ should be considered as lost \citep{mic02}.  

%--------------------
\begin{figure}[h]
\centering
\includegraphics[width=\columnwidth]{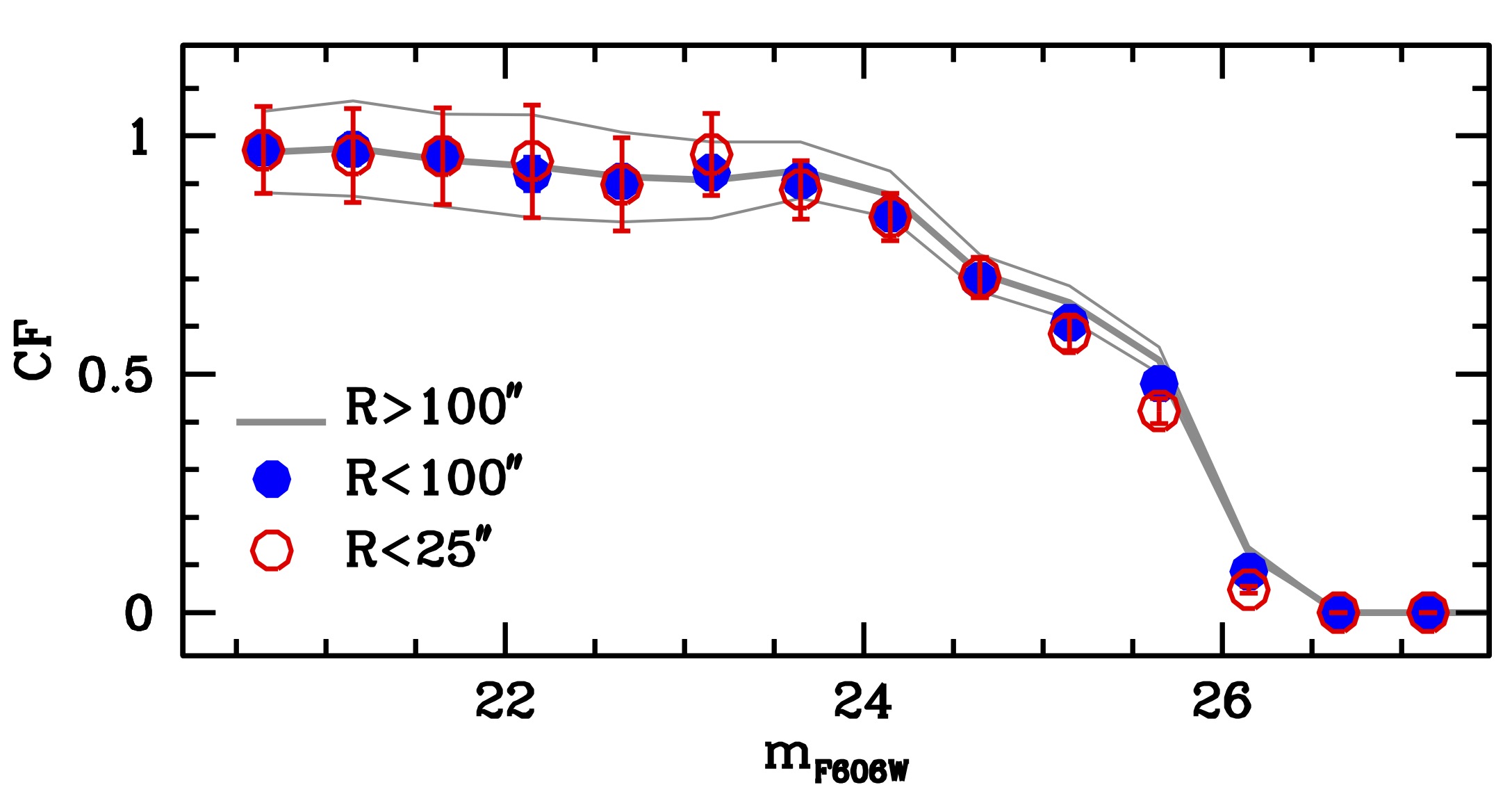}
\caption{Completeness Fraction (CF) as a function of magnitude as derived from artificial-star experiments for different radial regions of the ACS-WFC field. The CF for $R>100\arcsec$ is taken for reference, and is plotted as a continuous thick grey line, with the two thin lines enclosing the range of the uncertainties.} \label{fig:cf}
\end{figure}
%--------------------

In Fig.~\ref{fig:cf} we show that the CF does not depend on the distance of the star from the cluster centre, in the range of magnitudes that we used in the analysis performed in Sect.~\ref{sect:struc} and Sect.~\ref{sec:lf}. Small differences in the CF curves of different radial regions only appear for $\rm m_{F606W}>25.5$. Based on this evidence, the LF shown in Fig.~\ref{fig:lf} has been corrected according to the CF derived from the entire set of artificial stars in the ACS-WFC field. 

\section{Isochrone fitting}
\label{app:isofit}

\begin{figure}
\centering
\begin{minipage}{0.45\columnwidth}
\centering        
\includegraphics[width=0.9\columnwidth]{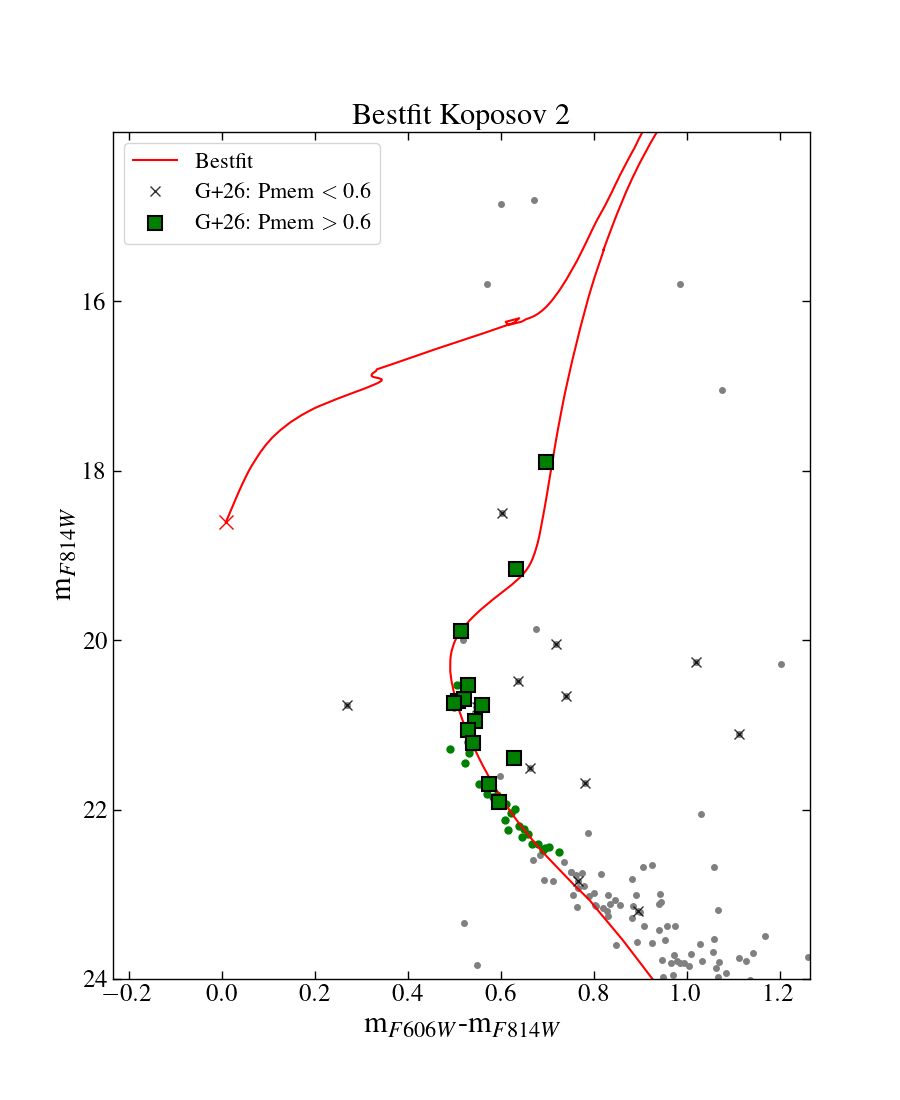}           
\end{minipage}
\begin{minipage}{0.45\columnwidth}
\centering
\includegraphics[width=0.9\columnwidth]{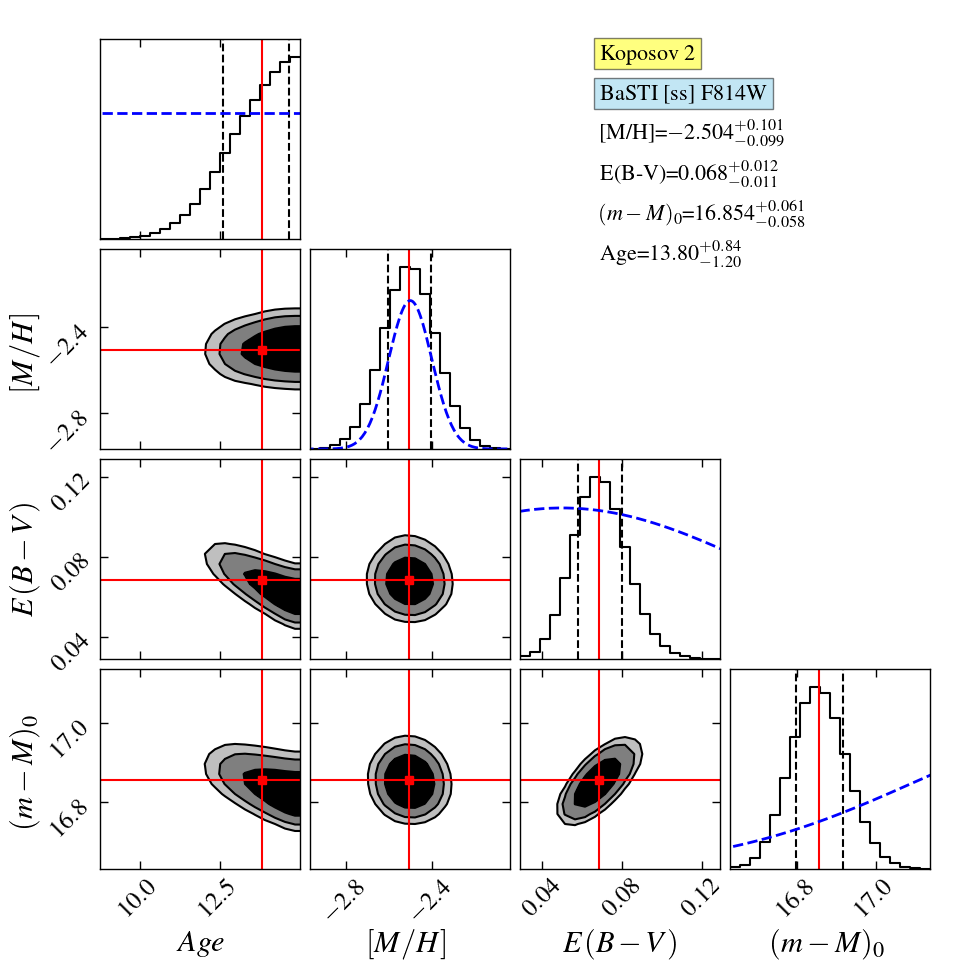}    
\end{minipage}  
\caption{Results of the Bayesian isochrone fit in the $m_{606W}-m_{814W}$ vs. $m_{814W}$ plane. The meaning of the symbols and the arrangement of the figure are the same as Fig.~\ref{fig:agedist} .}
\label{fig:age814}
\end{figure}

Since, following \citet{massari23,massari_25}, we compute the final values of the parameters derived from the isochrone fitting process
as the mean of the results obtained from the fit in the ${\rm m}_{\rm F606W}-{\rm m}_{\rm F814W}$ vs. ${\rm m}_{\rm F606W}$ and in the ${\rm m}_{\rm F606W}-{\rm m}_{\rm F814W}$ vs. ${\rm m}_{\rm F814W}$
planes, we show here, in Fig.~\ref{fig:age814} the fit in the CMD having $\rm m_{\rm F814W}$ on the $y$ axis.

\begin{figure}
\centering
\begin{minipage}{0.4\columnwidth}
\centering        
\includegraphics[width=0.9\columnwidth]{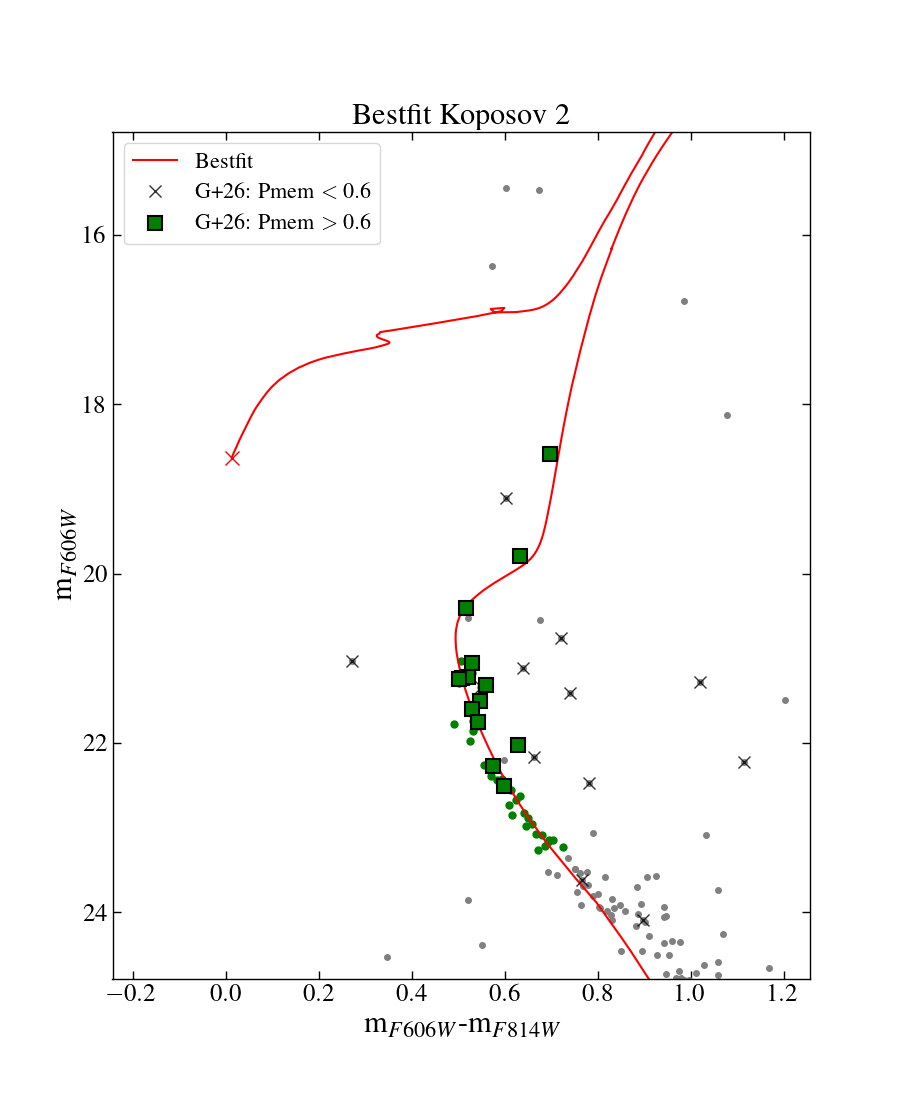}           
\end{minipage}
\begin{minipage}{0.4\columnwidth}
\centering
\includegraphics[width=0.9\columnwidth]{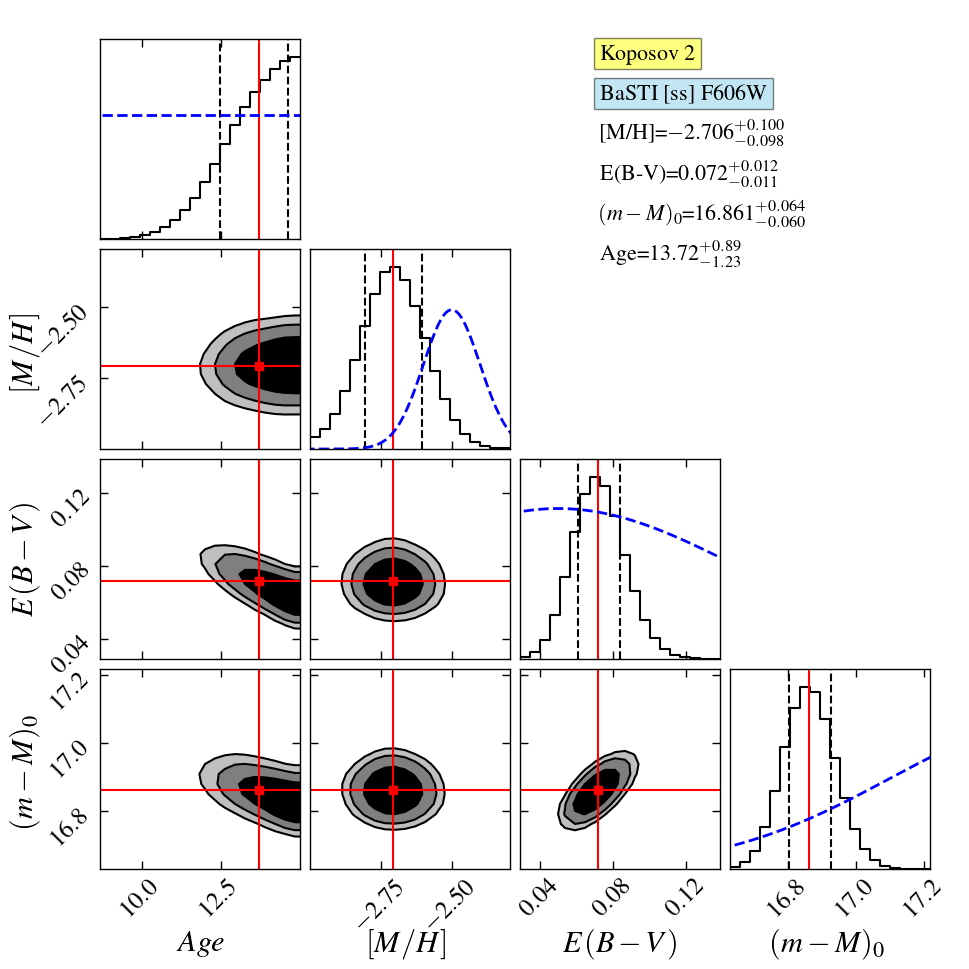}    
\end{minipage}  
\begin{minipage}{0.4\columnwidth}
\centering        
\includegraphics[width=1.0\columnwidth]{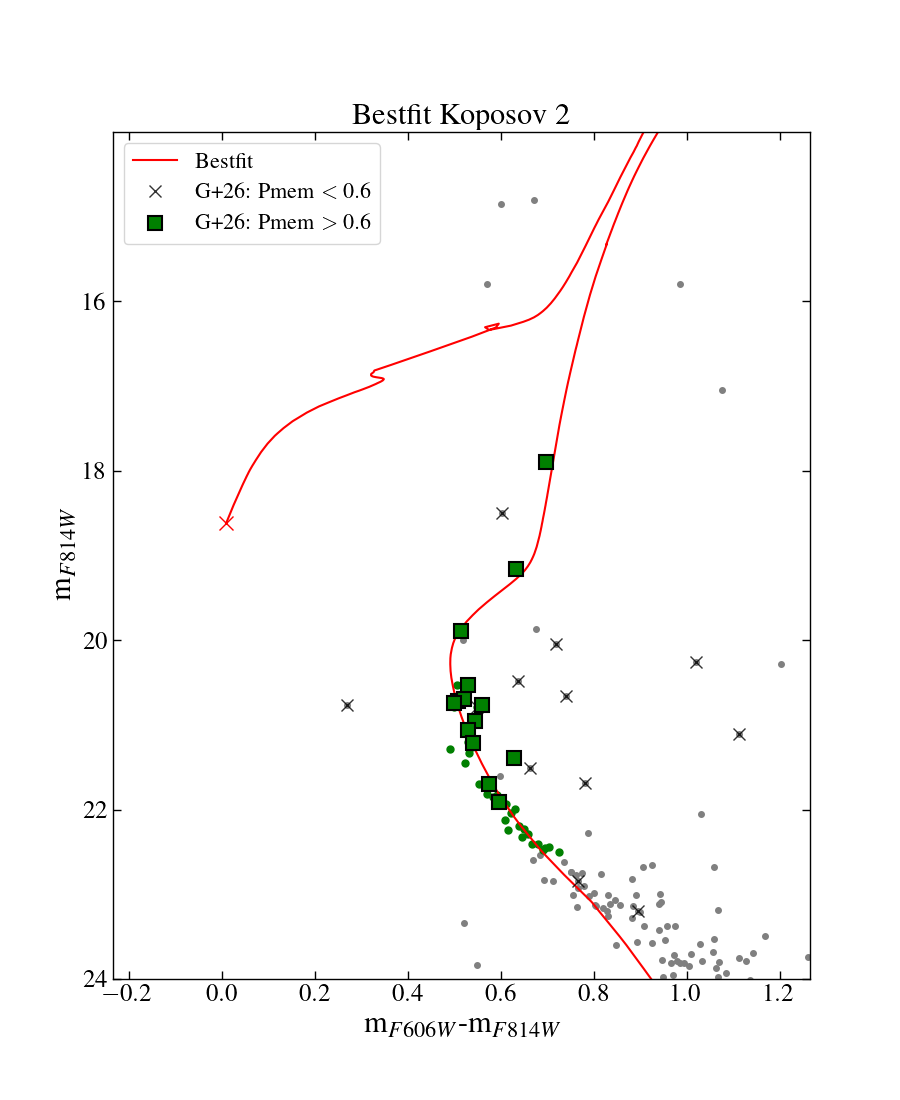}           
\end{minipage}
\begin{minipage}{0.4\columnwidth}
\centering
\includegraphics[width=1.0\columnwidth]{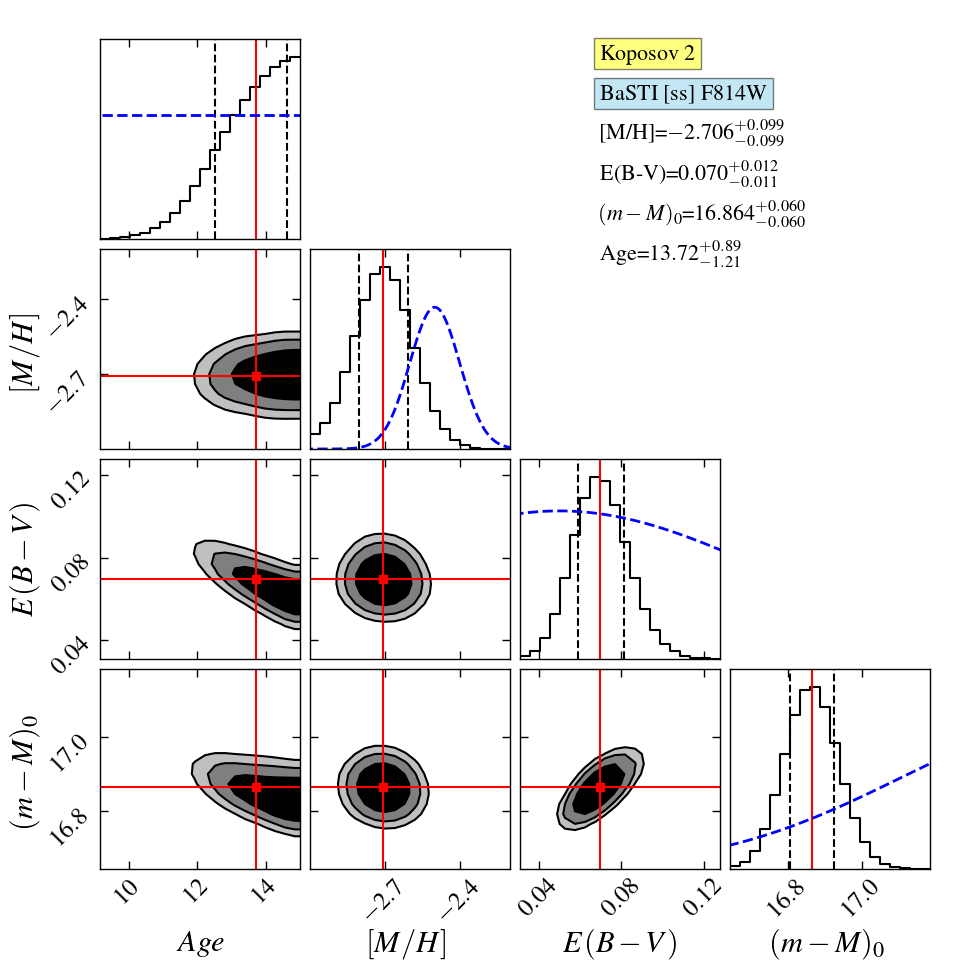}    
\end{minipage}  

\caption{Results of the Bayesian isochrone fit in the ${\rm m}_{\rm F606W}-{\rm m}_{\rm F814W}$ vs. ${\rm m}_{\rm F606W}$ plane (upper panels) and ${\rm m}_{\rm F606W}-{\rm m}_{\rm F814W}$ vs. ${\rm m}_{\rm F814W}$ plane (lower panels) with the same priors as in Fig.~\ref{fig:agedist} and Fig.~\ref{fig:age814} except for the gaussian prior on metallicity that here is centred on $\rm [M/H]=-2.7$, corresponding to $\rm [Fe/H]=-2.9$ and $\rm [\alpha/Fe]=+0.3$, instead of  $\rm[M/H]=-2.5$, corresponding to $\rm [Fe/H]=-2.9$ and $\rm [\alpha/Fe]=+0.5$, as in Fig.~\ref{fig:agedist}. The meaning of the symbols and the arrangement of the figure are the same as Fig.~\ref{fig:agedist}.}
\label{fig:agem27}
\end{figure}

In Fig.~\ref{fig:agem27} we show the same set of plots for an isochrone fit in which all the priors are the same as in Fig.~\ref{fig:agedist} and Fig.~\ref{fig:age814} except for the gaussian prior on metallicity that here is centred on $\rm [M/H]=-2.7$,
corresponding to $\rm [Fe/H]=-2.9$ and $\rm [\alpha/Fe]=+0.3$, instead of  $\rm[M/H]=-2.5$,
corresponding to $\rm [Fe/H]=-2.9$ and $\rm [\alpha/Fe]=+0.5$, as in Fig.~\ref{fig:agedist}. The resulting distance, age and reddening are indistinguishable from the case presented in Sect.~\ref{sect:agedist} as our baseline, showing that changes in $\rm [\alpha/Fe]$ within the range expected for metal-poor stars in globular clusters or dwarf galaxies do not affect our main conclusions.

%-------------------------------------------------------------
%-------------------------------------------------------------

%-------------------------------------------------------------
%-------------------------------------------------------------

\FloatBarrier

\section{Orbital parameters}
\label{app:orb}

Based on the present-day position and motion the orbit of Kop~2 was integrated backward using two different codes:
\begin{itemize}
    \item First, we integrated the cluster orbit using \texttt{AGAMA} \citep{vasiliev_19} in a Galactocentric reference frame, adopting the same position and velocities for the Sun as in \citet{ceccarelli_26}. We modelled the Milky Way potential following the prescriptions by \citet{mcmillan_17}. To account for observational uncertainties, we performed $100$ Monte Carlo orbit simulations, sampling the six-dimensional phase-space input parameters (see Table \ref{tab:input_op}) assuming Gaussian errors. The final orbital parameters were derived from the respective distributions by adopting their median values and corresponding $\pm1\sigma$ uncertainties.
    \item Then, we repeated the integration with the in-house code \textsc{OrbIT}, presented and fully described in  \citet{deleo26a,deleo26b}. Briefly, the \citet{deleo26a} potential is a modified form of \citet{mcmillan_17} potential with updated parameters for all components and a composite bulge made of a rotating bar and a spheroid. We kept the Solar phase-space values needed for the transformation to the Galactocentric reference frame equal to the previous integration and accounted for the observational uncertainties in the same way.
    \item Finally, we tested the inclusion of the Large Magellanic Cloud (LMC) as a perturber. We did this with a variant of the \textsc{OrbIT} code where the LMC is a point sourcing its potential, which is modelled following recent studies on the LMC structure \citep{jimarr24, watkins24}, with the geometry coming from the model with the LMC bar and disc aligned to the line-of-nodes of \citet{kacharov24}. Finally, the LMC values for its starting position and velocity are taken from Gaia EDR3 \citep{jimarr23}.
\end{itemize}

The orbital and dynamical parameters from the different integrations are reported in Table \ref{tab:op}. The most significant differences (in energy, orbital parameters and period) between the orbits obtained with \texttt{AGAMA} and those obtained with \textsc{OrbIT} should be attributed to the different Dark Matter halo mass of the underlying MW potential adopted, $1.3\times 10^{12}~M_{\sun}$ for the former, and $8.0\times 10^{11}~M_{\sun}$ for the latter, since the details of the shape of the various components of the potential only have a minor effect in the range of Galactocentric distances spanned by the orbit of Kop~2. The introduction of the LMC produces quite a few changes in the orbit of Kop~2, evidencing how the MW closest and biggest satellite is an important perturber of halo tracers.

%-----------------
\begin{table}[h]
\caption{Observables adopted for orbit integration.}
\label{tab:input_op}
\centering
\renewcommand{\arraystretch}{1.5}
\small
\setlength{\tabcolsep}{5.6pt}
\begin{tabular}{ccccccc}
\hline\hline 
Cluster & RA & Dec & $\mu_{\alpha}\rm cos(\delta)$ & $\mu_{\delta}$ & Distance & $V_{\rm los}$ \\
 & [deg]  & [deg] & [$\rm mas \; yr^{-1}$] & [$\rm mas \; yr^{-1}$] & [kpc] & [$\rm km \; s^{-1}$]\\
\hline
Koposov~2  & $119.571$ (1) & $26.255$ (1) & $-0.745\pm0.056$ (2) & $0.140\pm0.042$ (2) & $23.4\pm0.6$ (4) & $112\pm9$ (3) \\									    
\hline 
\end{tabular}
\tablefoot{Numbered references are: (1) \citet{vasiliev_21}, (2) \citet{libra26}, (3) \citet{ceccarelli_26}, and (4) this work.}
\vspace{2pt}
\end{table}
%-----------------

%-----------------
\begin{table*}[h]
\caption{Orbital parameters for Koposov~2.}
\label{tab:op}
\centering
\renewcommand{\arraystretch}{1.5}
\small
\setlength{\tabcolsep}{5.6pt}
\begin{tabular}{ccccccccc}
\hline\hline 
Potential & Energy & $L_{\rm z}$ & $L_{\rm perp}$ & $r_{\rm peri}$ & $r_{\rm apo}$ & $Z_{\rm max}$ & circ & P$_{\rm orb}$\\
 & [$10^{5} \; \rm km^{2} \; s^{-2}$]  & [$10^{3} \; \rm kpc \; km \; s^{-1}$] & [$10^{3} \; \rm kpc \; km \; s^{-1}$] & [kpc] & [kpc] & [kpc] &  &[Gyr] \\
\hline
McM17  & $-0.76^{+0.02}_{-0.02}$ & $7.86^{+0.29}_{-0.19}$ & $2.94^{+0.18}_{-0.18}$ & $28.7^{+0.7}_{-0.7}$ & $59.9^{+4.8}_{-3.4}$ & $20.3^{+1.4}_{-1.5}$ & $0.83^{+0.01}_{-0.02}$ & $0.96^{+0.04}_{-0.04}$\\									    
DL26      & $-0.37^{+0.02}_{-0.02}$ & $7.85^{+0.23}_{-0.23}$                 & $2.95^{+0.13}_{-0.13}$                 & $29.1^{+0.8}_{-0.8}$               & $101.8^{+9.8}_{-9.8} $      & $35.3^{+3.2}_{-3.2}$               & $0.88^{+0.03}_{-0.03}$  & $1.84^{+0.18}_{-0.18}$  \\
DL26+LMC & $-0.37^{+0.02}_{-0.02}$ & $7.85^{+0.23}_{-0.23}$                 & $2.95^{+0.13}_{-0.13}$  & $28.5^{+4.3}_{-4.3}$               & $91.9^{+11.2}_{-11.2}$    & $48.3^{+17.3}_{-17.3}$ & $0.88^{+0.03}_{-0.03}$   &  $1.62^{+0.18}_{-0.18}$ \\
\hline 
\end{tabular}
\vspace{2pt}
\tablefoot{McM17 = \citet{mcmillan_17}; DL26 = \citet{deleo26a}.
The orbital circularity (circ) is defined as the ratio of the cluster $L_z$ to the $L_z$ of a maximally rotating planar orbit with the same energy \citep{massari19, abadi+2003}.}
\end{table*}
%-----------------

\FloatBarrier

\section{N-body simulations}\label{app:evol}
%--------------------
\begin{figure}[!th]
\centering
\includegraphics[width=0.45\textwidth]{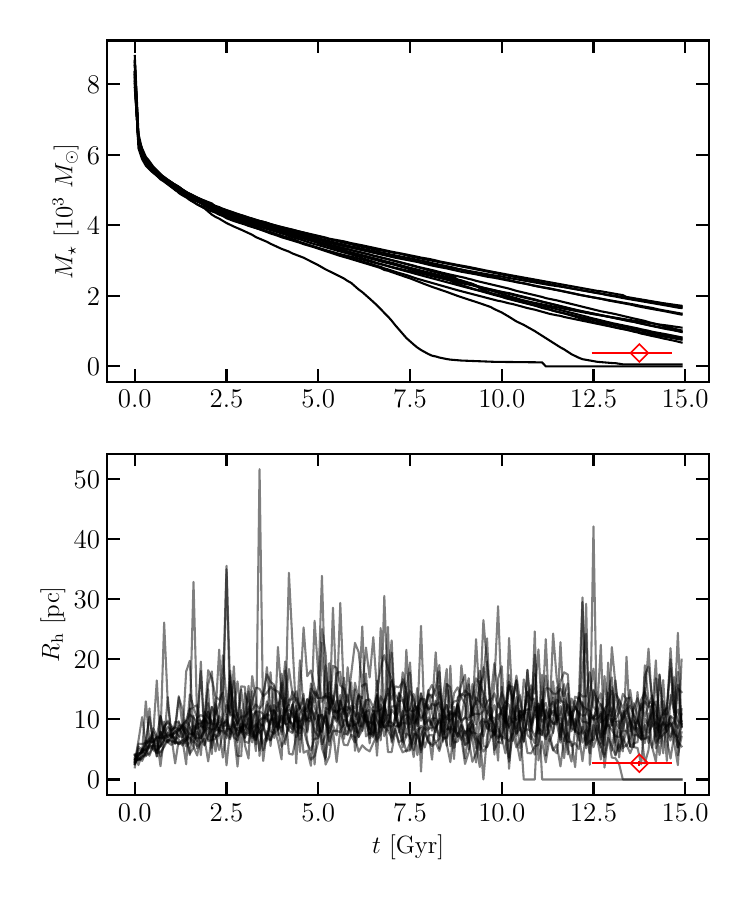}
\caption{
Evolution of a low-mass star cluster orbiting on a circular orbit at $R_{\rm GC}=30.0$~kpc from the Galactic center.
Different curves present the time evolution of the stellar mass (upper panel) and half-light radius (lower panel) for 15 different system realizations.
In red, we show the present-day properties of Kop~2.
}\label{fig:evol}
\end{figure}
\begin{figure}[!th]
    \centering
    \includegraphics[width=0.95\linewidth]{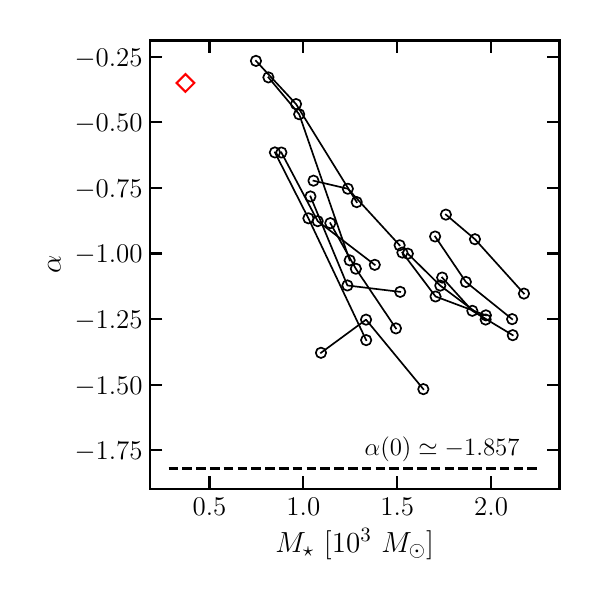}
    \caption{Evolution of the MF slope ($\alpha$) as a function of the residual mass derived for the mass range $[0.38; 0.72]~M_\odot$ (the same adopted for Kop~2, see Fig.~\ref{fig:lf}). For each simulation, we derived the MF slope at three different times (i.e., 12.4~Gyr, 13.7~Gyr, and 14.6~Gyr, with times increasing from right to left, corresponding to the age range inferred for Kop~2). The horizontal dashed line shows the primordial value from the IMF. Finally, the red point marks the position of Kop~2.
    }
    \label{fig:mf_evolution}
\end{figure}
%--------------------

In this section, we discuss the dynamical evolution of low-mass star clusters in weak tidal fields. The aim is neither to provide a model for Kop~2 nor to constrain its initial properties (e.g., mass or size), but rather to probe how low-mass stellar clusters at large Galactocentric distances could evolve into systems with observable properties similar to those of Kop~2.

We present a set of 15 direct $N$-body simulations performed with the \texttt{Petar} code \citep{wang_etal2020_petar}, following the evolution of a low-mass star cluster for 15 Gyr. Each simulation is a different realization of the same system, composed of 13000 particles that sample a \citet{kroupa_etal2001_imf} IMF (resulting in initial masses between $7.9-8.8\times 10^3~M_\odot$) and are initially distributed according to a \citet{king66} model with $W_0=7$ and half-mass radius of 4~pc.
We initialized each simulation on a circular orbit at 30~kpc in a \texttt{MWPotential2014} \citep{bovy_2015_galpy}. Although we acknowledge that Kop~2 is on a different orbit, $R_{\rm GC}=30$~kpc is similar to the peri-galactic radius of the system, and we stress that the simulations presented here are not intended to provide a model for the specific case.
Finally, we did not include any primordial binaries in the simulation, and for stellar evolution purposes, we assumed a metallicity of $Z=0.001$.

In Fig.~\ref{fig:evol}, we show the evolution of the total system mass within the Jacobi radius (eq.~\ref{for:jac}) and of the half-light radius for all the models.
After the initial mass loss and expansion driven by stellar evolution, the system slowly loses mass due to two-body relaxation. Interestingly, some realizations dissolve much earlier than 15~Gyr, probably due to stochastic IMF sampling. Systems that survive for 13.7~Gyr or longer have masses between $1000-2000~M_\odot$.
The system's half-light radius is about $\simeq6-11~$pc, although with large fluctuations due to the dearth of stars. 
{ The line-of-sight velocity dispersion at the cluster centre is always $<0.5$ \kms, hence the surviving clusters remain dynamically  cold even if they are approaching disruption.}

{ We also measured the $r$ parameter (as defined in \citealt{baum22}) for all the simulations that survive until 13.7~Gyr. We adopted the same magnitude ranges as done in Sect.~\ref{sec:lf} for the observations. Across the 15 different realizations, we measure values of $r$ in the range $0.84^{+0.05}_{-0.04} - 1.02^{+0.04}_{-0.03}$, suggesting little or no mass segregation and consistent with what we measured in the real system in Sect.~\ref{sec:lf}.}

In addition, in Fig.~\ref{fig:mf_evolution} we show the evolution of the mass function slope at late times (from 12.4~Gyr to 14.6~Gyr) as a function of the present-day mass. Due to two-body relaxation, the mass function flattens over time \citep[see e.g.,][]{vh97, bm03, trenti_2010}, reaching values broadly consistent with those observed in Kop~2. The evaporation of stars through the tidal boundary would thus provide a simple explanation for a flattened present-day mass function. 
We argue that if the system were embedded in a massive dark matter halo, it would be much harder for stars to escape and for the system to flatten its MF by losing a significant fraction of its stars.

{ It may be interesting to point out that the simulations consistently include stellar remnants. We computed the fraction of mass in white dwarfs (WD) and in black holes (BH) at the end state of the surviving models: $\rm M_{WD}/M_{TOT}$ ranges from 37\% to 43\%, while $\rm M_{BH}/M_{TOT}<$3\%. Neutron stars are virtually always ejected from the system by the natal kick in low mass clusters as those considered here, hence do not contribute to the mass budget at old ages.}

The small set of simulations presented here highlights the relevance of internal dynamical evolution and mass loss for low-mass systems even at very large Galactocentric distances (i.e., weak tidal fields). 
Their present-day properties are broadly consistent with those of UFCSs \citep[][such as Kop~2, see Fig.~\ref{fig:evol}]{cerny_26}.
In addition, depending on the specific initial configuration, some of these systems might be dissolving, making them observable only at large distances from the Galactic center, where dissolution times are long enough.
A systematic exploration of different initial conditions (in terms of mass, size, and orbital properties) and the comparison with the known population of UFCSs \citep{cerny_26} will be the subject of a future, tailored study.

\end{appendix}

\end{document}